\documentclass{svmult}

\usepackage{graphicx}  %
\usepackage{amsmath}
\usepackage{bm}  %

\newcommand{\bea}{\begin{eqnarray}}
\newcommand{\eea}{\end{eqnarray}}
\newcommand{\beq}{\begin{equation}}
\newcommand{\eeq}{\end{equation}}
\newcommand{\bqa}{\begin{eqnarray}}
\newcommand{\eqa}{\end{eqnarray}}

\def\mqo2{{\!\!\!}}

\begin{document}

\author{~}

\date{\today}
\title*{Universal Relations for Fermions \\
      with Large Scattering Length}

\maketitle

{\large

\noindent
{\bf Eric Braaten}\\
Department of Physics,
The Ohio State University, Columbus, OH\ 43210, USA

\vfill

\noindent
{\bf Abstract}\\
The behavior of fermions with two spin states that interact 
with a large scattering length is constrained by 
universal relations that hold for any state of the system.
These relations involve a central property of the system 
called the contact, which measures the number of pairs 
of fermions with different spins that have small separations.
The contact controls the thermodynamics of the system 
as well as the large-momentum and high-frequency tails 
of correlation functions. 
This review summarizes the current theoretical and 
experimental status of these universal relations.

\vfill

\noindent
\rule{\linewidth}{.7pt}\\
to be published as a chapter in 
{\bf BCS-BEC Crossover and the Unitary Fermi Gas} 
(Lecture Notes in Physics),
edited by Wilhelm Zwerger (Springer, 2011).
}

\newpage

\section{Introduction}
\label{sec:introduction}

Particles with short-range interactions that produce a 
large scattering length have universal properties that
depend only on the scattering length \cite{Braaten:2004rn}.
A system consisting of such particles is
strongly interacting in the sense that there are effects of the
interactions that must be treated nonperturbatively.  
These strong interactions give rise to strong correlations 
among the particless.  Many theoretical methods,
even if they are nonperturbative, are inadequate 
for dealing with such strong correlations.
However, such a system is also governed by {\it universal relations} 
that follow from the short-distance and short-time dynamics
associated with the large scattering length.
These universal relations provide powerful constraints 
on the behavior of the system.  They
hold for any state of the system: few-body or
many-body, ground state or nonzero temperature, homogeneous or in a
trapping potential, normal state or superfluid, balanced in the two
spin states or imbalanced.  They connect various properties of the
system, ranging from thermodynamic variables to large-momentum and
high-frequency tails of correlation functions.

The systems for which the universal relations have been most extensively 
studied are those consisting of fermions with two spin states.
The universal relations that have been derived thus far all involve a
property of the system called the {\it contact}, which measures the
number of pairs of fermions in the two spin states with small
separations.  Many of these relation were first derived by Shina Tan,
and they are known as the 
{\it Tan relations}~\cite{Tan:0505,Tan:0508,Tan:0803}.  
Tan derived these relations by exploiting the fact that 
the large scattering length can be taken into account through boundary
conditions on the many-body Schr\"odinger wavefunction
for otherwise noninteracting particles.  
The universal relations can also be derived
concisely within a quantum field theory framework
\cite{Braaten:2008uh}, where they follow from renormalization
and from the operator product expansion.  
Such a framework facilitates the derivation of additional universal
relations and the systematic inclusion of corrections associated with
the nonzero range of the interactions.

In this review, we summarize the current theoretical and experimental 
status of universal relations for systems consisting of fermions 
with two spin states and a large scattering length.
We begin in Section~\ref{sec:tan-relations} by presenting the Tan relations. 
In Section~\ref{sec:what-contact}, we discuss the 
physical interpretation of the contact 
and we provide some illustrative examples. 
In Section~\ref{sec:other}, we present other universal relations 
that have been derived more recently.
In Section~\ref{sec:experiment}, we describe exciting recent developments 
in the field of ultracold atoms involving measurements of the 
contact and experimental tests of the universal relations.
In Section~\ref{sec:deriv-univ-results}, we discuss the derivation
of the universal relations, with an emphasis on 
quantum field theory methods. 

\section{The Tan Relations}
\label{sec:tan-relations}

The Tan relations were derived by Shina Tan 
in a series of three papers~\cite{Tan:0505,Tan:0508,Tan:0803}. 
The first two of these papers were written in 2005,
but they were not published until 2008,
when all three papers were published back-to-back in
Annals of Physics. 

The Tan relations apply to systems consisting of fermions 
with two spin states 
whose scattering length $a$ is large compared to
the range $r_0$ of their interactions.
We will refer to the fermions as atoms 
and label the two spin states by an index $\sigma$ with values 1 and 2. 
In a many-body system, the number densities $n_\sigma$ 
and the temperature $T$ must also be small enough that the 
corresponding length scales are large compared to the range:  
$n_\sigma^{-1/3} \gg r_0$ and $\lambda_T \gg r_0$, 
where $\lambda_T = (2 \pi \hbar^2/m k_B T)^{1/2}$.
If the system is in an external trapping potential
$V(\bm{r}) = \frac12 m \omega^2 r^2$, the length scale associated 
with the trap should also be large compared to the range:
$(\hbar/m \omega)^{1/2} \gg r_0$.

The Tan relations involve an extensive
quantity, the {\it contact} $C$, which is the integral over space 
of a local quantity, the
{\it contact density} ${\cal C}(\bm{R})$:
\begin{equation}
C = \int \!\! \hbox{d}^3R~  {\cal C}(\bm{R}) .
\label{C-Cdensity}
\end{equation}
We proceed to present
the Tan relations in chronological order.

\subsection{Tails of distributions}

In the first of Tan's two 2005 papers, he derived three universal relations
\cite{Tan:0505}.  The first was for the tails of the momentum
distributions $n_{\sigma}(\bm{k})$ for the two spin states 
$\sigma = 1,2$:
\begin{description}
\item 
{\bf Tail of the momentum distribution}.
The distributions of the wavevector $\bm{k}$
have power-law tails at large $k$:
\begin{equation}
n_{\sigma} (\bm {k}) \longrightarrow C/k^4.
\label{tails}
\end{equation}
The coefficient $C$ is the contact and it is the same 
for both spin states.  
\end{description}
The asymptotic behavior in Eq.~(\ref{tails}) 
actually applies only in the scaling region 
$|a|^{-1} \ll k \ll r_0^{-1}$.  The wavenumber must also 
be larger than the scales associated with the system,
such as $n^{1/3}$, $\lambda_T^{-1}$, and $(m \omega/\hbar)^{1/2}$.
The momentum distributions in Eq.~(\ref{tails})
have been normalized so that 
the total number of atoms in the spin state $\sigma$ is
\begin{equation}
N_\sigma = \int \!\! \frac{\hbox{d}^3k}{(2 \pi)^3} ~ n_{\sigma}(\bm {k}). 
\end{equation}
The universal relation in Eq.~(\ref{tails}) implies that
the contact is positive definite and has dimension (length)$^{-1}$.
Thus the contact density has dimensions (length)$^{-4}$.

The total energy $E$ of the system is the sum of the 
kinetic energy $T$, the interaction energy $U$, 
and the energy $V$ associated with an external potential: 
\begin{equation}
E = T + U + V.
\label{E-TUV}
\end{equation}
The kinetic energy $T$  
(which should not be confused with the temperature)
can be expressed as an integral over the momentum distribution:
\begin{equation}
T \equiv 
\sum_\sigma \int \!\! \frac{\hbox{d}^3k}{(2 \pi)^3} 
\left( \frac{\hbar^2 k^2}{2m} \right)
        n_{\sigma}(\bm{k}).
\label{T-div}
\end{equation}
The asymptotic behavior 
of the momentum distribution in Eq.~(\ref{tails}) 
implies that $T$ is ultraviolet divergent.
This divergence actually occurs only in the zero-range limit
$r_0 \to 0$.  For interactions with a finite range, 
the integral in Eq.~(\ref{T-div})
is cut off by the range and therefore has
a contribution that behaves like $1/r_0$ as $r_0 \to 0$.
Thus the physical interpretation of the ultraviolet divergence
is that $T$ is sensitive to the range.
The second Tan relation in Ref.~\cite{Tan:0505} implies that the 
sensitivity of the kinetic energy to the range is cancelled 
by the interaction energy:
\begin{description}
\item
{\bf Energy relation}.
The sum of the kinetic and interaction energies 
is ultraviolet finite and it is
completely determined by the momentum distributions 
$n_{\sigma}(\bm{k})$ and the contact $C$:
\begin{equation}
T + U = 
\sum_\sigma \int \!\! \frac{\hbox{d}^3k}{(2 \pi)^3} \frac{\hbar^2 k^2}{2m} 
        \left( n_{\sigma}(\bm{k}) - \frac{C}{k^4} \right)
+ \frac{\hbar^2}{4 \pi m a} C . 
\label{energy}
\end{equation}
\end{description}
In the integral on the right side, the subtraction term 
cancels the tail of the momentum distribution 
and makes the integral convergent in the ultraviolet.
The sum of the two terms in Eq.~(\ref{energy}) 
proportional to the contact $C$ is the interaction energy.
The first of those two terms is ultraviolet divergent.
Thus the interaction energy is sensitive to the range, but
that sensitivity is exactly cancelled by the kinetic energy.
The last term in Eq.~(\ref{energy}) is the interaction 
energy that remains after subtracting the divergent term.
Remarkably, it is also proportional to the contact.

The third Tan relation in Ref.~\cite{Tan:0505} gives the 
asymptotic behavior of the correlation function for the 
densities of the two spin states at short distances:
\begin{description}
\item
{\bf Density-density correlator at short distances}.
The correlation between the number densities for the two spin states
at points separated by a small distance $r$ diverges as $1/r^2$ 
and the coefficient of the divergence 
is proportional to the contact density:
\begin{equation}
\left \langle n_1(\bm{R} + \mbox{$\frac{1}{2}$}\bm{r})~
n_2(\bm{R} - \mbox{$\frac{1}{2}$} \bm{r}) \right \rangle
\longrightarrow \frac{1}{16 \pi^2}  
\left( \frac{1}{r^2} - \frac{2}{a r} \right) {\cal C}(\bm{R}).
\label{nncor}
\end{equation}
\end{description}
Tan also pointed out that the contact density appears in the 
short-distance expansion for the correlator of the 
quantum field operators that create and annihilate the atoms.
This expansion will be discussed in Section~\ref{sec:short-dist-OPE}.

\subsection{Changes in the scattering length}

From the three universal relations described above,
one might conclude that the contact is an esoteric property of the system
that has only to do with tails of distributions.
In the second of Tan's 2005 papers~\cite{Tan:0508},
he derived another universal relation that makes it clear that
the contact is an absolutely central property of the system:
\begin{description}
\item
{\bf Adiabatic relation}.
The rate of change of the energy 
due to a small change in the inverse scattering length is
proportional to the contact: 
\begin{equation}
\left( \frac {\hbox{d} E~~}{\hbox{d} a^{-1}} \right)_S = 
- \frac{\hbar^2}{4\pi m} ~ C.
\label{adiabatic}
\end{equation}
The derivative is evaluated with the entropy $S$ held fixed.
The particle numbers $N_1$ and $N_2$ are also implicitly held fixed.
\end{description}
In the simplest case, $E$ is just an energy eigenvalue.
The adiabatic relation also holds for any 
statistical mixture of eigenstates if the derivative is 
evaluated with the occupation numbers held fixed.
By the adiabatic theorem of quantum mechanics,
if the scattering length changes sufficiently slowly
with time, the occupation numbers remain constant.
Thus if the contact $C$ is known as a function of $a$,
the adiabatic relation in Eq.~(\ref{adiabatic}) can be integrated 
to obtain the accumulated change in $E$.

The adiabatic relation can also be expressed in terms of the derivative 
of the free energy $F = E - T S$ with the temperature $T$ 
held fixed:
\begin{equation}
\left( \frac {\hbox{d} F~~}{\hbox{d} a^{-1}} \right)_T = 
- \frac{\hbar^2}{4\pi m} ~ C.
\label{adiabatic2}
\end{equation}
As pointed out by Tan, this implies that the contact 
determines the thermodynamics of the system.  
Given the contact of a system as a function of the scattering length $a$
and other variables, such as $N_1$, $N_2$, and $T$, 
the free energy $F$ can be obtained by integrating Eq.~(\ref{adiabatic2})
with respect to $a$.  A convenient boundary condition is 
provided by the limit $a \to 0^-$, in which the atoms are
noninteracting.  From $F$, one can determine all the other 
thermodynamic functions.

If one uses Eq.~(\ref{tails}) to define the contact in terms 
of the tail of the momentum distribution, 
this appears to be a case of the tail wagging the dog.
The thermodynamic behavior of the system seems to be determined 
by the tail of the momentum distribution.
However the proper interpretation is that the contact 
is a central property of the system that
determines both the thermodynamics 
and the tail of the momentum distribution.

The adiabatic relation in Eq.~(\ref{adiabatic}) determines
the change in the total energy when the scattering length
changes very slowly.  Tan also considered the opposite limit 
in which the scattering length changes very rapidly~\cite{Tan:0508}: 
\begin{description}
\item
{\bf Sudden change in the scattering length}.
If the scattering length is changed suddenly from $a$ to $a'$,
the change in the total energy is proportional to the contact:
\begin{equation}
\Delta E = 
-\frac{\hbar^2}{4\pi m}\left( \frac{1}{a'} - \frac{1}{a} \right)~ C,
\label{sudden}
\end{equation}
where $C$ is the initial value of the contact.
\end{description}
This result requires the time scale for the sudden change in 
scattering length to be much slower than the time scale
$m r_0^2/\hbar$ associated with the range.
Tan also presented a more general result for the change in the energy 
due to a rapid change in the scattering length,
which will be described in Section~\ref{sec:rapid}.

\subsection{Additional Tan relations}

In Tan's 2008 paper, he derived two additional 
universal relations that apply for specific forms of the 
external potential~\cite{Tan:0803}.
\begin{description}
\item
{\bf Virial theorem}.
For a system in a harmonic trapping potential, 
the components of the energy $E$ in Eq.~(\ref{E-TUV}) satisfy
\begin{equation}
T + U - V = - \frac{\hbar^2}{8 \pi m a} ~ C.
\label{virial}
\end{equation}
\end{description}
The virial theorem for the unitary limit $a = \pm \infty$ 
was first derived and also verified experimentally 
by Thomas, Kinast, and Turlapov~\cite{TKT0504}.
The virial theorem in Eq.~(\ref {virial}) is the generalization
to finite scattering length.
\begin{description}
\item
{\bf Pressure relation}.
For a homogeneous system, 
the pressure and the energy density are related by
\begin{equation}
{\cal P} = \frac{2}{3}{\cal E} + \frac{\hbar^2}{12\pi m a} ~  \cal C.
\label{pressure}
\end{equation}
\end{description}
The pressure relation was actually first derived in Ref.~\cite{Tan:0508}
for the special case of a balanced gas in which the two spin states 
have equal populations.  The derivation was extended to the 
general case in Ref.~\cite{Tan:0803}.

If there are inelastic 2-body 
scattering processes with a large energy release, 
they will result in a decrease in the number of low-energy atoms.
Tan realized that the rate at which the number density
of low-energy atoms decreases is proportional 
to the contact density ${\cal C}$~\cite{Tan-private}.
The proportionality constant was first given in 
Ref.~\cite{Braaten:2008uh}.
If there are inelastic 2-body scattering channels,
the scattering length $a$ has a negative imaginary part.
The proportionality constant in the universal relation 
can be expressed in terms of that 
complex scattering length:
\begin{description}
\item
{\bf Inelastic 2-body losses}.
If there are inelastic 2-body scattering processes with a 
large energy release, the number density of low-energy atoms 
decreases at a rate that is proportional to the contact density:
\begin{equation}
\frac{\hbox{d} ~}{\hbox{d} t} n_\sigma(\bm{R}) = 
-  \frac{\hbar (-{\rm Im}\,a)}{2 \pi m  |a|^2} ~ {\cal C}(\bm{R}).
\end{equation}
\end{description}

\section{What is the Contact?}
\label{sec:what-contact}

Given the Tan relations described in Section~\ref{sec:tan-relations}, 
it is evident that the contact is a central property of the system.
But what is it?  In this section, we provide an intuitive interpretation 
of the contact.  We also provide additional insights into 
the contact by giving analytic expressions in some simple cases.

\subsection{Intuitive interpretation}

An intuitive interpretation of the contact density can be derived 
from the universal relation for the density-density 
correlator in Eq.~(\ref{nncor}).  That relation can be expressed 
in the form
\begin{equation}
\left \langle n_1(\bm{R} + \bm{r}_1)~
n_2(\bm{R} + \bm{r}_2) \right \rangle
\longrightarrow \frac{1}{16 \pi^2 |\bm{r}_1 - \bm{r}_2|^2} {\cal C}(\bm{R}).
\label{nncor2}
\end{equation}
If we integrate both $\bm{r}_1$ and $\bm{r}_2$ over a ball of radius $s$,
we obtain
\begin{equation}
N_{\rm pair}(\bm{R},  s)  \longrightarrow
\frac{s^4}{4} {\cal C}(\bm{R}).
\label{Npair}
\end{equation}
The left side simply counts the number of pairs inside that ball,
which is the product $N_1 N_2$ of the number of atoms in
the two spin states.  The volume of that ball is $V = \frac43 \pi s^3$.
One might naively expect the number of pairs to scale as $V^2$
as $V \to 0$. However, according to Eq.~(\ref{Npair}),
it scales instead as $V^{4/3}$.
That scaling behavior applies only for $s$ smaller than $|a|$ 
and also smaller than the length scales associated with the system, 
such as $n^{-1/3}$, $\lambda_T$, and $(\hbar/m \omega)^{1/2}$.
The scaling behavior extends down to $s$ of order the range $r_0$.  

A naive definition of the density of pairs is the limit as $V \to 0$ 
of $N_1 N_2/V^2$, where $N_1$ and $N_2$ are the numbers of 
atoms in the volume $V$.  
This quantity has dimensions (length)$^{-6}$.
The result in Eq.~(\ref{Npair}) implies that the combination with a
nontrivial small-volume limit is $N_1 N_2/V^{4/3}$,
which has dimensions (length)$^{-4}$.
Thus a more appropriate definition of the local pair density is
the small-volume limit of $N_1 N_2/V^{4/3}$, 
up to a normalization constant that can be chosen by convention.
The unusual dimensions of (length)$^{-4}$ for this local pair density
can be expressed concisely by saying that 
this quantity has {\it scaling dimension} 4.  
The difference $-2$ between the scaling dimension 
and the naive dimension 6 is called the {\it anomalous dimension}.
This anomalous dimension comes from the strong correlations 
associated with the large scattering length.
This anomalous scaling behavior implies that the number of pairs 
in a very small volume is much larger than one would naively expect
by extrapolating from larger volumes.
The contact density is a measure of the local pair density
that takes into account this anomalous scaling behavior.

Further intuition for the contact can be gleaned from the universal 
relation for the tail of the momentum distribution in Eq.~(\ref{tails}).
It implies that the number of atoms in either spin state
whose wavenumber $\bm{k}$ is larger than $K$ is
\begin{equation}
N_\sigma(|\bm{k}| > K) = \frac{1}{2 \pi^2 K} C ,
\label{N-CK}
\end{equation}
provided $K$ is in the scaling region $1/|a| \ll K \ll 1/r_0$
and is larger than the wavenumber scales set by the system.
Thus the contact is a measure of the number of atoms with large momentum.

When interpreting the contact density as a measure of the 
local pair density, one should not confuse those pairs with 
Cooper pairs, which are pairs with a specific momentum correlation.
Under conditions in which Cooper pairs are well defined,
the typical separation of the atoms in a Cooper pair is much larger than 
the interparticle spacing. 
The number of Cooper pairs in a volume $V$ 
that is comparable to or smaller than the volume per particle
is not well defined. 
It certainly does not have the anomalous scaling behavior 
$V^{-4/3}$ of the total number of pairs.

\subsection{Few-body systems}

The adiabatic relation in Eq.~(\ref{adiabatic}) can be used as an operational
definition of the contact.  If the energy of a system is known as a
function of the scattering length, we can simply differentiate 
to get an expression for the contact.
A simple example in the case $a > 0$ is the
weakly-bound diatomic molecule, or {\it shallow dimer}, consisting of two
atoms with spins 1 and 2.  The universal result for its 
binding energy is $\hbar^2/m a^2$.  Its energy is therefore
\begin{equation}
E_\textrm{dimer} = - \frac{\hbar^2}{m a^2}.
\label{E-dimer}
\end{equation}
Using the adiabatic relation in Eq.~(\ref{adiabatic}),
we find that the contact for the dimer is 
\begin{equation}
C_\textrm{dimer} = \frac{8 \pi}{a}.
\label{C-dimer}
\end{equation}

Blume and Daily have calculated the contact numerically 
for the ground state of 4 fermions, 2 in each spin state,
trapped in a harmonic potential and interacting through a potential 
with a large adjustable scattering length $a$~\cite{BD:0909}.  
They determined the contact 
as a function of $a$ using four different universal
relations:  the tail of the momentum distribution in 
Eq.~(\ref{tails}), the density-density correlator at short distances 
in Eq.~(\ref{nncor}), the adiabatic relation in Eq.~(\ref{adiabatic}),
and the virial theorem in Eq.~(\ref{virial}).
The small differences between the four determinations of the contact
were compatible with systematic errors associated 
with the nonzero range of the potential.

\subsection{Balanced homogeneous gas}

The contact density ${\cal C}$ for the homogeneous gas
can be obtained by using the adiabatic relation in 
Eq.~(\ref{adiabatic}) as an operational definition.
Dividing both sides of the equation by the volume, 
the relation can be expressed in the form
\begin{equation}
{\cal C} = \frac{4\pi m a^2}{\hbar^2}~\frac {\hbox{d} {\cal E}}{\hbox{d} a}.
\label{adiabatic-C}
\end{equation}
Analytic results for the energy density ${\cal E}$ are available 
in various limits, and they can be used to obtain analytic 
expressions for the contact density.

We first consider the case of a balanced gas, in which the two 
spin states are equally populated, at zero temperature.
The total number density $n = 2 n_1 = 2 n_2$ determines 
the Fermi momentum:  $k_F = (3 \pi^2 n)^{1/3}$.  
The ground state is determined by the dimensionless 
interaction variable $1/k_F a$.  As this variable changes from 
$- \infty$ to 0 to $+\infty$, the ground state changes smoothly 
from a mixture of two weakly-interacting Fermi gases 
to a Bose-Einstein condensate of diatomic molecules.
The ground state is always a superfluid.
The mechanism for superfluidity evolves smoothly from the 
BCS mechanism, which is the Cooper pairing of atoms 
in the two spin states, to the BEC mechanism, which is the  
Bose-Einstein condensation of dimers.

In the {\it BCS limit}  $a \to 0^-$, 
the energy density can be expanded in powers of $k_F a$:
\begin{equation}
{\cal E} =
\frac{\hbar^2 k_F^5}{10 \pi^2 m} 
\left(1 + \frac{10}{9 \pi} k_F a + \ldots \right) .  
\end{equation}
Using Eq.~(\ref{adiabatic-C}), we find that the contact density 
in the BCS limit is
\begin{equation}
{\cal C} \longrightarrow 4 \pi^2 n^2 a^2.  
\label{C-BCS}
\end{equation}
This is proportional to the low-energy cross section $4 \pi a^2$ and 
to the product $(n/2)^2$ of the number densities of the two spin states.
The contact density decreases to 0 as $a \to 0^-$, 
but it decreases only as $a^2$.  
This result emphasizes that the contact density
is not closely related to the density of Cooper pairs, 
which decreases exponentially as $\exp(- \pi/k_F|a|)$
as $a \to 0^-$.

In the {\it unitary limit}  $a \to \pm \infty$, 
the energy density can be expanded in powers of $1/k_F a$:
\begin{equation}
{\cal E} = \frac{\hbar^2 k_F^5}{10 \pi^2 m} 
\left( \xi  - \frac{\zeta}{k_F a} + \ldots\right) ,
\end{equation}
where $\xi$ and $\zeta$ are numerical constants.
Using Eq.~(\ref{adiabatic-C}), we find that the contact density 
in the unitary limit is
\begin{equation}
{\cal C} \longrightarrow \frac{2 \zeta}{5 \pi} (3 \pi^2 n)^{4/3}.
\label{C-unitary}
\end{equation}
Since the interaction provides no length scale in the unitary limit,
the contact density must be proportional to $k_F^4$ 
by dimensional analysis.
An estimate $\zeta \approx 1$ for the numerical constant in
Eq.~(\ref{C-unitary})
can be obtained from numerical calculations of the energy density 
near the unitary limit using quantum Monte Carlo methods 
\cite{CPCS:0406,ABCG:0406}.
A more precise value can be obtained from numerical calculations 
of the density-density correlator in the unitary limit
using the fixed-node diffusion Monte Carlo method~\cite{LCGRS:0604}.
Using the Tan relation for the density-density correlator
in Eq.~(\ref{nncor}), we obtain the value
\begin{equation}
\zeta \approx 0.95.
\label{zeta}
\end{equation}

In the {\it BEC limit}  $a \to 0^+$, 
the energy density can be expanded in powers of $(k_F a)^{3/2}$:
\begin{equation}
{\cal E} =  - \frac{\hbar^2 n}{2 m a^2} 
+ \frac{\pi \hbar^2 n^2 a_{dd}}{4m} 
\left( 1 + \frac{128}{15} \sqrt{n a_{dd}^3/2 \pi} + \ldots \right),
\end{equation}
where $a_{dd}\approx 0.60~a$ is the 
dimer-dimer scattering length~\cite{PSS:0309}. 
The leading term is the total binding energy density
for dimers with number density $n/2$ and binding energy given by
Eq.~(\ref{E-dimer}).
The second term is the mean-field energy of a Bose-Einstein
condensate of dimers with dimer-dimer scattering length $a_{dd}$.
Using Eq.~(\ref{adiabatic-C}), we find that 
the contact density in the BEC limit is
\begin{equation}
{\cal C} \longrightarrow \frac{4 \pi n}{a}.
\label{C-BEC}
\end{equation}
This is equal to the contact $8 \pi/a$ for a dimer,
which is given in Eq.~(\ref{C-dimer}),
multiplied by the dimer number density $n/2$.
The contact density diverges as $1/a$ as $a \to 0^+$.
The first correction to the leading term in Eq.~(\ref{C-BEC})
is suppressed by a factor of $(k_F a)^3$.

\begin{figure}[t]
\centerline{\includegraphics*[width=4.in,angle=0]{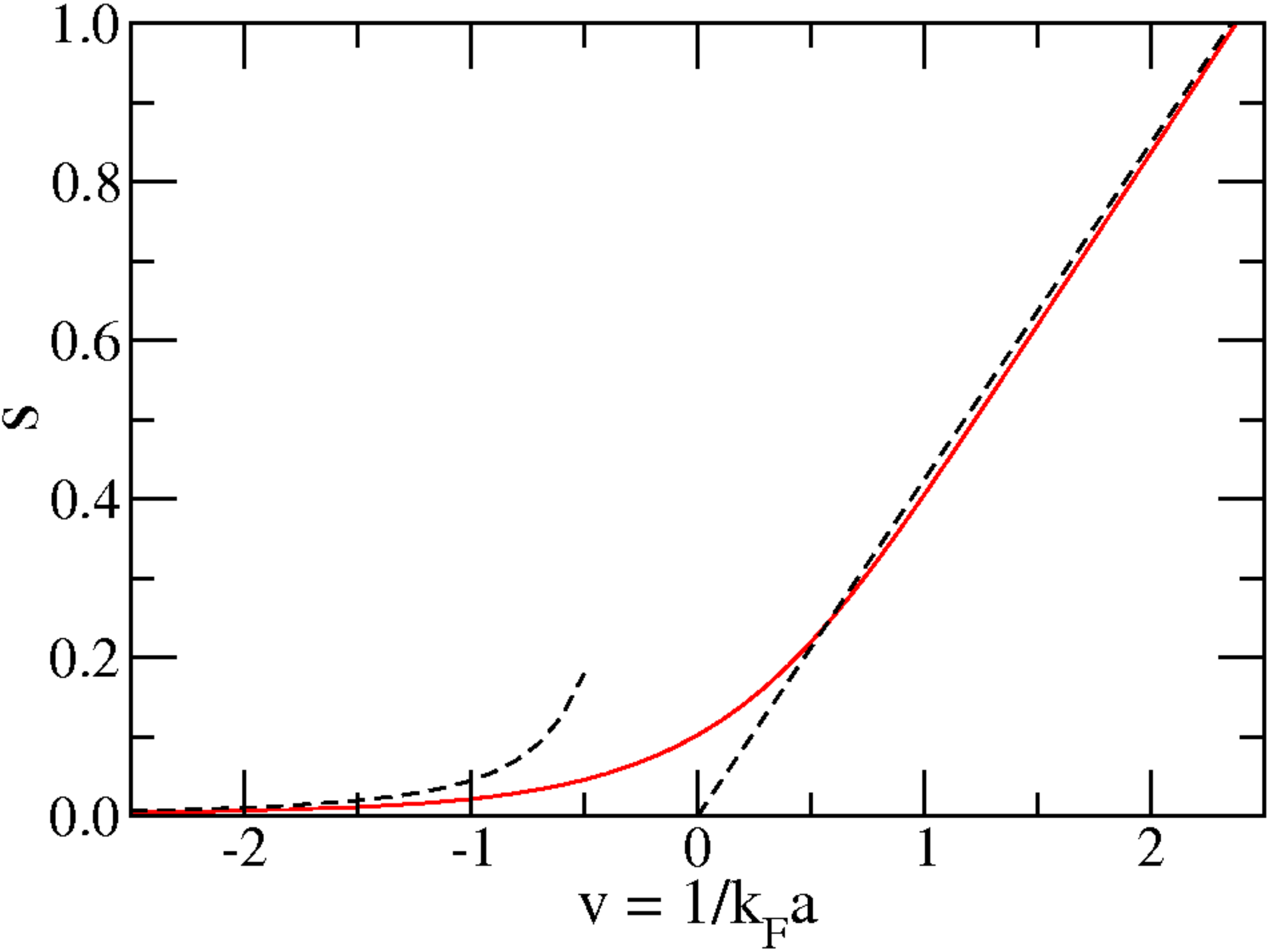}}
\caption{
The dimensionless contact density $s = {\cal C}/k_F^4$ 
for the balanced homogeneous gas as a function of 
the dimensionless coupling strength $1/k_F a$,
from Ref.~\cite{HPZ:0904}.
The left dashed line is the leading contribution
in the BCS limit given by Eq.~(\ref{C-BCS}).
The right dashed line is the leading contribution
in the BEC limit given by Eq.~(\ref{C-BEC}).
The contact density in the unitary limit may be 
underpredicted by about 16\%.
}
\label{fig:balanced}
\end{figure}
The contact density ${\cal C}$ for the ground state of the 
balanced homogeneous gas can be expressed as $k_F^4$
multiplied by a monotonically-increasing dimensionless function 
of $1/k_F a$.
Haussmann, Punk, and Zwerger have calculated 
this function numerically~\cite{HPZ:0904}.
They used the Luttinger-Ward self-consistent formalism
to calculate the single particle spectral functions. 
The contact density was determined using the Tan relation 
in Eq.~(\ref{tails}) for the tail of the momentum distribution.   
Their result for the contact density
as a function of $1/k_F a$ is shown in Fig.~\ref{fig:balanced}.
It interpolates smoothly between the BCS
limit in Eq.~(\ref{C-BCS}) and the BEC limit in Eq.~(\ref{C-BEC}). 
Their result in the unitary limit corresponds to a value 
$\zeta \approx 0.80$ for the constant in Eq.~(\ref{C-unitary}).
This is about 16\% smaller than the value in Eq.~(\ref{zeta})
that was obtained from a fixed-node diffusion Monte Carlo
calculation.  
This difference reflects a systematic theoretical error
in the calculational method of Ref.~\cite{HPZ:0904}.

The dependence of the contact density 
on the temperature has been determined analytically in 
various limits by Yu, Bruun, and Baym~\cite{YBB:0905}.
In the low-temperature limit, the leading thermal contribution 
to the contact density comes from phonons.
The leading thermal correction at small $T$ 
increases like $T^4$~\cite{YBB:0905}.
It becomes significant when $T$ is 
comparable to the Fermi temperature: $k_B T_F = \hbar^2 k_F^2/2m$.
In the BCS limit and in the unitary limit,
the thermal contribution differs from the contact at $T=0$
by a factor of $(T/T_F)^4$ multiplied by a numerical constant.
In the BEC limit, the thermal contribution is suppressed 
by a factor of $(T/T_F)^4$ and by a further factor of $(k_F a)^{1/2}$.

The contact density at high temperature can be calculated 
using the virial expansion. 
The high-temperature limit\footnote{
In Ref.~\cite{YBB:0905}, the contact density was denoted by 
$16 \pi^2 \hbar^2 C$ and the total number density
was denoted by $2n$ instead of $n$.} 
was determined by Yu, Bruun, and Baym~\cite{YBB:0905}:
\begin{equation}
{\cal C} \longrightarrow \frac{8 \pi^2 \hbar^2 n^2}{m k_B T}.  
\end{equation}
Since the contact density increases as $T^4$ at small $T$
and decreases as $1/T$ at sufficiently large $T$, 
it must achieve a maximum somewhere in between.
The maximum occurs for $T$ of order $T_F$.
The maximum is pronounced only when $a$ is 
near the unitary limit~\cite{YBB:0905}.

Palestini, Perali, Pieri, and Strinati have calculated 
the contact density for the balanced homogeneous gas
at nonzero temperature using a
diagrammatic t-matrix approximation~\cite{PPPS:1005}.
They determined the contact from the high-momentum tail of the 
momentum distribution in Eq.~(\ref{tails})
and from the high-frequency tail of the 
radio-frequency transition rate (which is discussed in 
Section~\ref{sec:rf}). 

Thus far, the calculational methods that have been used to 
calculate the contact density numerically
involve uncontrolled approximations.
While they may be accurate in certain limits, 
there may also be regions of $k_F a$ and $T/T_F$
in which the systematic theoretical errors are not negligible.

\subsection{Strongly-imbalanced homogeneous gas}

We now consider the strongly-imbalanced gas, 
in which a tiny population of minority atoms in state 2 
is immersed in a system of atoms in state 1.
The minority atoms can be considered as a dilute gas of impurities 
in the Fermi sea of majority atoms.
In the homogeneous gas with number densities $n_1$ and $n_2$,
the ground state is determined by the 
dimensionless interaction variable $1/k_{F1} a$,
where $k_{F1} = (6 \pi^2 n_1)^{1/3}$ is the Fermi wavenumber 
for the majority atoms.
In the BCS limit $a \to 0^-$, the impurity particle is an atom in state 2.
In the BEC limit $a \to 0^+$, the impurity particle is the dimer
whose binding energy is given by Eq.~(\ref{E-dimer}).
Using a diagrammatic Monte Carlo method, Prokof'ev and Svistunov 
have shown that there is a phase transition at a critical value 
$a_c$ of the scattering length given by  
$1/k_{F1} a_c = 0.90 \pm 0.02$~\cite{PS:0707}.
As $1/k_{F1} a$ increases through this critical value, the impurity 
changes from a quasiparticle associated with the atom in state 2, 
which is called a {\it polaron}, to a dimer quasiparticle.

Analytic expressions for the contact density for the 
ground state of the strongly-imbalanced homogeneous gas
can be obtained from the energy density using Eq.~(\ref{adiabatic-C}).
In the {\it BCS limit} $a \to 0^-$, the contact density is
\begin{equation}
{\cal C} \longrightarrow 16 \pi^2 n_1 n_2 a^2.
\label{C2-BCS}
\end{equation}
This is proportional to the low-energy cross section $4 \pi a^2$ and 
to the product $n_1 n_2$ of the two number densities.
The contact density is an increasing function of $1/a$
that is smooth except for a discontinuity 
at the phase transition at $a=a_c$.
In the {\it BEC limit}  $a \to 0^+$, the contact density is 
\begin{equation}
{\cal C} \longrightarrow \frac{8 \pi n_2}{a}.
\label{C2-BEC}
\end{equation}
This is equal to the contact $8 \pi/a$ for the dimer,
which is given in Eq.~(\ref{C-dimer}),
multiplied by the dimer number density $n_2$. 

\begin{figure}[t]
\centerline{\includegraphics*[width=4.in,angle=0]{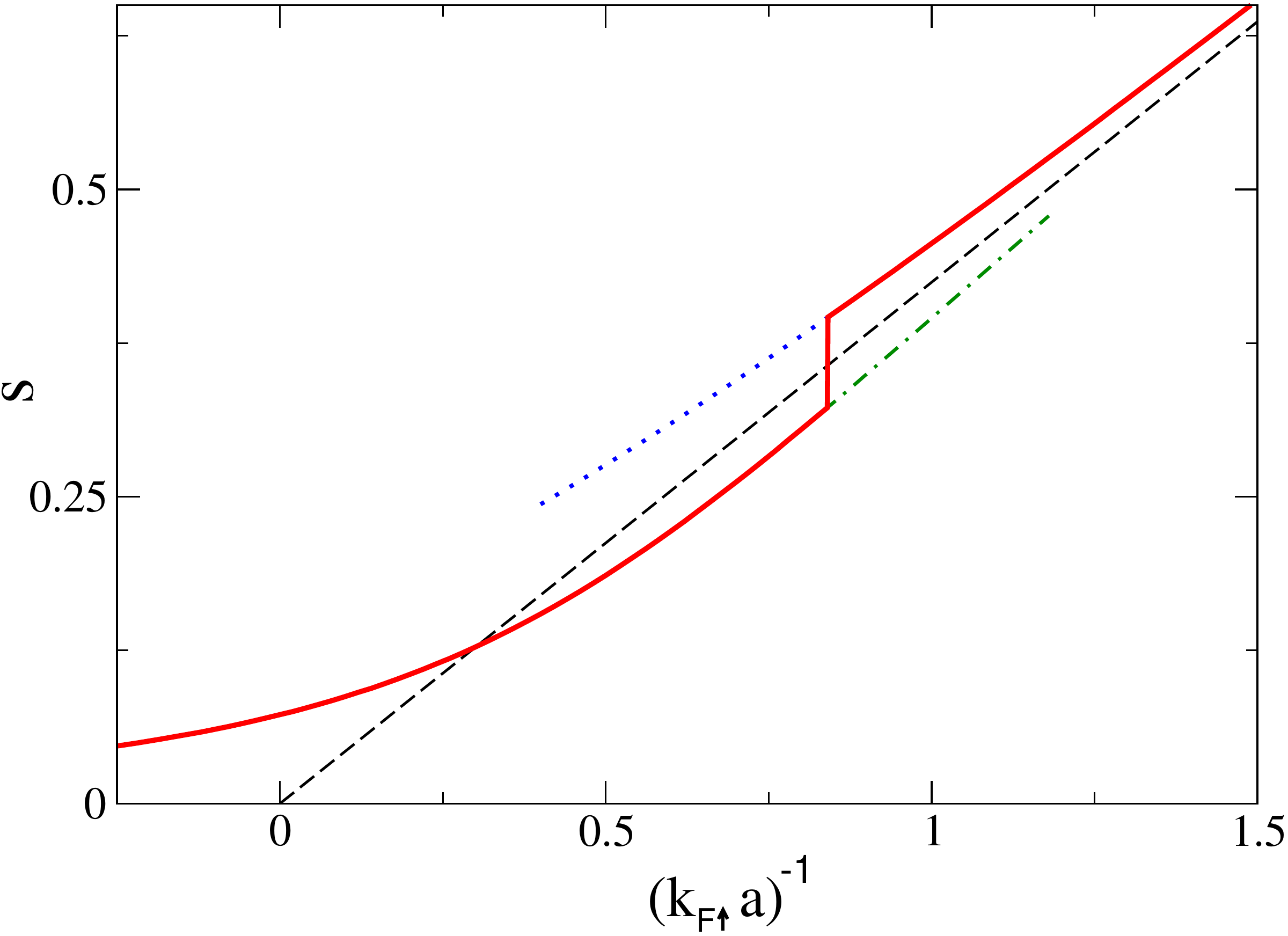}}
\caption{
The dimensionless contact density $s = {\cal C}/(6 \pi^2 k_{F1} n_2)$ 
for the strongly-imbalanced homogeneous gas as a function of 
the dimensionless coupling strength $1/k_{F1} a$,
from Ref.~\cite{PDZ:0908}.
The dashed line is the leading contribution
in the BEC limit given by Eq.~(\ref{C2-BEC}).
The dotted and dash-dotted lines are continuations of the 
solid line past the phase transition into metastable regions. 
The position of the phase transition 
may be underpredicted by about 7\%.
}
\label{fig:imbalanced}
\end{figure}
The contact density ${\cal C}$ for the ground state of the
strongly-imbalanced homogeneous gas can be expressed as 
$k_{F1} n_2$ multiplied by a monotonically-increasing 
dimensionless function of $1/k_{F1} a$.
This function has been calculated numerically by 
Punk, Dumitrescu, and Zwerger~\cite{PDZ:0908}.
They used the adiabatic relation in Eq.~(\ref{adiabatic-C})
to obtain the contact from the ground state energy density,
which they calculated using a variational method 
that gives a fairly good approximation 
to the results from the diagrammatic Monte Carlo method~\cite{PS:0707}.
The results of Ref.~\cite{PDZ:0908} for the contact density 
as a function of $1/k_{F1} a$
are shown in Fig.~\ref{fig:imbalanced}.
For large negative values of  $1/k_{F1} a$, the contact density 
can be approximated by the BCS limit in Eq.~(\ref{C2-BCS}).
It is predicted to increase to about $5~k_{F1} n_2$ 
in the unitary limit $a \to \pm \infty$
and then to about $20~k_{F1} n_2$ at the phase transition $a_c$,
where it is predicted to have a discontinuity of about 20\%. 
The prediction of the variational method for the position 
of the phase transition is $(k_{F1} a_c)^{-1} \approx 0.84$, 
which is about 7\% smaller than the
diagrammatic Monte Carlo result $0.90 \pm 0.02$.
This difference reflects a systematic theoretical error
in the variational method of Ref.~\cite{HPZ:0904}.
For large positive values of $1/k_{F1} a$, the contact density 
can be approximated by the BEC limit in Eq.~(\ref{C2-BEC}).

\section{Other Universal Relations}
\label{sec:other}

Many new universal relations involving the contact have been 
discovered in recent years.  They reveal that the contact plays 
a central role in many of the most important probes for ultracold atoms.

\subsection{rf spectroscopy}
\label{sec:rf}

Given a system of atoms in spin states 1 and 2,
a radio-frequency (rf) signal that is tuned to near the transition 
frequency between an atom in state 2 and an atom in a third
spin state 3 can transform the atoms in state 2 coherently 
into linear combinations of atoms in states 2 and 3.  
These atoms can be subsequently transformed 
by decohering processes into a mixture of atoms in states 2 and 3.
The net effect is a transfer of atoms from the state 2 to the state 3.
The transition rate $\Gamma(\omega)$ for this process depends 
on the frequency $\omega$ of the rf signal.  It is convenient to choose 
the offset for $\omega$ to be the transition frequency for a single
atom.  The transition rate for an extremely dilute sample of $N_2$ atoms
is then a delta function at $\omega = 0$ :
\begin{equation}
\Gamma(\omega) \longrightarrow
\pi \Omega^2~\delta(\omega)~N_2,
\label{eq:Gamma-N2}
\end{equation}
where $\Omega$ is the Rabi frequency associated with the transition.
In a many-body system consisting of atoms in states 1 and 2,
$\Gamma(\omega)$ can be modified by initial-state interactions
between atoms in states 1 and 2 and by final-state
interactions between atoms in state 3 and 
atoms in states 1 or 2.
However the effects of these interactions are constrained 
by a sum rule~\cite{YB:0510,ZL:0710}:
\begin{equation}
\int_{-\infty}^{\infty} \!\! \hbox{d}\omega~\Gamma(\omega) =
\pi \Omega^2~N_2.
\label{eq:Gamma-sum0}
\end{equation}

If the atoms interact through large pair scattering lengths
$a_{12} \equiv a$, $a_{13}$, and $a_{23}$, there are universal relations 
that govern the rf transition rate $\Gamma(\omega)$.
One of these universal relations is a sum rule derived by 
Punk and Zwerger~\cite{PZ:0707} and
by Baym, Pethick, Yu, and Zwierlein~\cite{BPYZ:0707}:
\begin{equation}
\int_{-\infty}^{\infty} \!\! \hbox{d}\omega~\omega \Gamma(\omega) =
\frac{\hbar \Omega^2}{4 m} \left( \frac{1}{a_{12}} - \frac{1}{a_{13}} \right) 
C_{12},
\label{eq:Gamma-sum1}
\end{equation}
where $C_{12} \equiv C$ is the contact for atoms in states 1 and 2.
The term proportional to $1/a_{13}$ comes from final-state 
interactions between atoms in states 1 and 3.
If we divide the sum rule in Eq.~(\ref{eq:Gamma-sum1})
by the sum rule in Eq.~(\ref{eq:Gamma-sum0}), we get an expression 
for the frequency shift $\langle \omega \rangle$
averaged over the system. 
This frequency shift is called the {\it clock shift}.
The universal relation for the clock shift has several 
interesting features.
If $a_{13} = a_{12}$, the clock shift vanishes 
because of a symmetry relating atoms 2 and 3.
The clock shift 
has smooth behavior in the unitary limit $a_{12} \to \pm \infty$.
This behavior was first observed in experiments 
on rf spectroscopy in $^6$Li atoms~\cite{ZHGK:0306}. 
If we take the limit $a_{13} \to 0$ in Eq.~(\ref{eq:Gamma-sum1}),
the sum rule diverges. 
This implies that if the scattering length
$a_{13}$ is not large, the clock shift is sensitive to the range.

Another universal relation for rf spectroscopy is that the 
high-frequency tail of $\Gamma(\omega)$ is proportional to the contact.
The general result for large scattering lengths
$a_{12}$ and $a_{13}$ was derived in Ref.~\cite{Braaten:2010dv}:
\begin{equation}
\Gamma(\omega)  \longrightarrow
\frac{\Omega^2 (a_{13}^{-1} - a_{12}^{-1})^2}
    {4 \pi \omega (m \omega/\hbar)^{1/2} (a_{13}^{-2} + m \omega/\hbar)}
~C_{12}.
\label{eq:Gamma-high}
\end{equation}
The asymptotic behavior in Eq.~(\ref{eq:Gamma-high}) holds
when $\omega$ is much larger than the many-body frequency scales
$\hbar k_F^2/m$ and $k_B T/\hbar$, but still much smaller than the 
frequency scale $\hbar/m r_0^2$ associated with the range.
If $\omega \gg \hbar/m a_{13}^2$,  
the high-frequency tail decreases as $\omega^{-5/2}$.
The result if the scattering length
$a_{13}$ is not large can be obtained by taking
the limit $a_{13} \to 0$ in Eq.~(\ref{eq:Gamma-high}): 
\begin{equation}
\Gamma(\omega)  \longrightarrow
\frac{\Omega^2}
    {4 \pi \omega (m \omega/\hbar)^{1/2}}
~C_{12}.
\label{eq:Gamma-high2}
\end{equation}
In this case, the high-frequency tail decreases as $\omega^{-3/2}$.
This scaling behavior was derived in Ref.~\cite{SSR:0903} and 
the coefficient was first calculated correctly
by Schneider and Randeria~\cite{SR0:910}.
The scaling behavior was also pointed out in Ref.~\cite{PPS:0811}. 
If $\Gamma(\omega)$ decreases asymptotically as $\omega^{-3/2}$,
the sum rule in Eq.~(\ref{eq:Gamma-sum1}) diverges.
Thus it is the high-frequency tail in Eq.~(\ref{eq:Gamma-high2})
that makes this sum rule sensitive to the range
in the case of $a_{13}$ that is not large.

\subsection{Photoassociation}

Photoassociation uses a laser to transfer pairs of low-energy atoms 
into an excited molecular state with very high energy.  
The wavefunction of the molecule has support only 
over very short distances much smaller than the range $r_0$,
so the pair of atoms must be very close together to have
a reasonable probability of making the transition.
If there is a closed-channel molecule near the 2-atom threshold 
that can be excited by the laser, it can dominate the 
photoassociation rate.  The rate is then proportional to the 
number $N_\textrm{mol}$ of closed-channel molecules:
\begin{equation}
\Gamma  =
\frac{\Omega^2}{\gamma}~N_\textrm{mol},
\label{eq:Gamma-Nmol}
\end{equation}
where $\Omega$ is the Rabi frequency of the laser 
and $\gamma$ is the line width of the excited molecule.

Werner, Tarruel, and Castin~\cite{WTC:0807} and
Zhang and Leggett~\cite{Zhang:0809} pointed out that 
if the large scattering length $a$ comes from the tuning of the
magnetic field $B$ to near a Feshbach resonance
associated with this closed-channel molecule,
then $N_\textrm{mol}$ is proportional to the contact.
If the zero of energy is chosen to coincide with the threshold for the atoms, 
the rate of change of the energy of the system 
with respect to the magnetic field can be expressed as
\begin{equation}
\frac{\hbox{d} E}{\hbox{d} B} = - \mu_\textrm{mol}~N_\textrm{mol},
\label{eq:dE-Nmol}
\end{equation}
where $\mu_\textrm{mol}$ is the difference between the 
magnetic moment of the closed-channel molecule and twice the
magnetic moment of an atom in the open channel.
The scattering length $a(B)$ near a Feshbach resonance at $B_0$
can be parametrized by
\begin{equation}
a(B) = a_\textrm{bg} \left( 1 - \frac{\Delta}{B - B_0} \right).
\label{eq:aFeshbach}
\end{equation}
Combining Eqs.~(\ref{eq:dE-Nmol}) and (\ref{eq:aFeshbach})
and using the adiabatic relation in Eq.~(\ref{adiabatic}),
we obtain an expression for the number of closed-channel 
molecules that is proportional to the contact~\cite{WTC:0807}:
\begin{equation}
N_\textrm{mol} = 
\frac{R_* \Delta^2}{4 \pi [\Delta - (B - B_0)]^2} C ,
\label{eq:Nmol-C}
\end{equation}
where $R_*$ is a positive length that characterizes 
the width of the Feshbach resonance:  
\begin{equation}
R_* = - \frac{\hbar^2}{m \mu_\textrm{mol} a_\textrm{bg} \Delta}.
\label{eq:R*}
\end{equation}
That length can also be expressed as 
$R_* = - \frac12 r_s$, where $r_s$ is the effective range 
at the center of the resonance $B=B_0$.

The universal relation between the number of closed-channel molecules
and the contact was previously derived formally by 
Braaten, Kang and Platter~\cite{Braaten:2008bi}, 
but they did not make the connection to photoassociation.

\subsection{Structure factors}

Structure factors encode information about density-density correlations
in a system.  For a many-body system of fermions with two spin states,
the correlations between the densities of the two spin states 
are particularly important.  The corresponding static structure factor
$S_{12}(q)$ for a homogeneous system is the 
Fourier transform in the separation vector $\bm{r}_1 - \bm{r}_2$
of the correlator $\langle n_1(\bm{r}_1) n_2(\bm{r}_2) \rangle$
of the two densities. The dynamic structure factor
$S_{12}(q,\omega)$ is the Fourier transform in 
the separation vector and the time interval 
between the two densities.  It encodes information about the degrees of 
freedom that can be excited by density fluctuations.
The static structure factor can be obtained by integrating 
$S_{12}(q,\omega)$ over $\omega$. 

If the scattering length is large, the static structure factor $S_{12}(q)$ 
has a high-momentum tail that decreases like $1/q$~\cite{CGS:0512}.
Hu, Liu, and Drummond have pointed out that this tail 
is proportional to the contact density~\cite{HLD:1003}:
\begin{equation}
S_{12}(q)
\longrightarrow \frac{1}{8}  
\left( \frac{1}{q} - \frac{4}{\pi a q^2} \right) {\cal C}.
\label{S12-tail}
\end{equation}
The normalization of $S_{12}(q)$ in Ref.~\cite{HLD:1003} differs 
from that in Eq.~(\ref{S12-tail}) by a factor of $2/N$, 
where $N = N_1 + N_2$ is the total number of atoms. 
The universal relation in Eq.~(\ref{S12-tail})
follows simply by taking the Fourier transform of the Tan relation 
in Eq.~(\ref{nncor}) for the density-density correlation at short distances.

Son and Thompson have studied the dynamic structure factor 
$S_{12}(q,\omega)$ in the unitary limit~\cite{Son:1002}. 
They showed that the leading contribution 
in the scaling limit $\omega \to \infty$ and $q \to \infty$
with $x = \hbar q^2/2 m \omega$ fixed is proportional to the contact density.
The coefficient of ${\cal C}$ is $(m \omega^3/\hbar)^{-1/2}$
multiplied by a complicated function of the dimensionless scaling variable
$x$, which they calculated analytically.
For small $x$, their result reduces to
\begin{equation}
S_{12}(q, \omega) \longrightarrow 
\frac{4 q^4}{45 \pi^2 \omega (m \omega/\hbar)^{5/2}}~{\cal C}.
\label{S12-tail2}
\end{equation}
Taylor and Randeria have also determined the high-frequency tail 
of the dynamic structure factor~\cite{T-R:2010}.
Their result for the limits $q \to 0$ followed by $\omega \to \infty$ 
differs from that in Eq.~(\ref{S12-tail2}) by a 
factor of 3/2.

\subsection{Viscosity spectral functions}
\label{sec:viscosity}

Taylor and Randeria have derived universal relations for the viscosity
spectral functions of a homogeneous gas~\cite{T-R:2010}.
They found that the shear viscosity spectral function $\eta(\omega)$
has a high-frequency tail that is proportional to the contact density:
\begin{equation}
\eta(\omega) \longrightarrow
\frac{\hbar^2}{10 \pi (m \omega/\hbar)^{1/2}}~{\cal C}.
\label{eta-tail}
\end{equation}
They also derived a sum rule for $\eta(\omega)$:
\begin{equation}
\int_0^\infty \!\! \hbox{d} \omega
\left( \eta(\omega) 
- \frac{\hbar^2 \cal C}{10\pi (m \omega/\hbar)^{1/2}} \right) =
\frac{\pi \hbar}{3} 
\left( {\cal E} - \frac{3 \hbar^2}{10 \pi m a}~{\cal C} \right),
\label{eta-sum}
\end{equation}
where ${\cal E}$ is the energy density.
Given the high-frequency tail of $\eta(\omega)$ in Eq.~(\ref{eta-tail}),
the subtraction term on the left side of Eq.~(\ref{eta-sum})
is necessary to make the integral convergent.
Enss, Haussmann, and Zwerger obtained a result for the high-frequency
tail of $\eta(\omega)$ that also decreases as $\omega^{-1/2}$
but has a coefficient smaller than that in Eq.~(\ref{eta-tail})
by a factor of 2/3~\cite{EHS:1008}.

Taylor and Randeria have also derived a sum rule 
for the bulk viscosity spectral function $\zeta(\omega)$~\cite{T-R:2010}:
\begin{equation}
\int_0^\infty \!\! \hbox{d}\omega~\zeta(\omega) =
\frac{\hbar^3}{72 m a^2} 
\left( \frac{\hbox{d} {\cal C}~~~}{\hbox{d} a^{-1}} \right)_{S/N} ,
\label{zeta-sum}
\end{equation}
where the derivative is taken with the entropy per particle held fixed.
Since the spectral function $\zeta(\omega)$ is positive definite,
the sum rule in Eq.~(\ref{zeta-sum}) implies that the contact density 
${\cal C}$ is a strictly increasing function of $a^{-1}$.

\subsection{Rapid change in the scattering length}
\label{sec:rapid}

In Ref.~\cite{Tan:0508}, Tan showed that if the scattering length $a(t)$ 
is time dependent, the total energy of the system changes 
at a rate that is proportional to the contact:
\begin{equation}
\frac{\hbox{d}~}{\hbox{d} t} E(t) = 
-\frac{\hbar^2}{4\pi m} C(t) \dot b(t),
\label{sweep}
\end{equation}
where $b(t) = 1/a(t)$ and  $C(t)$ is the instantaneous contact at time $t$.
If the external potential $V(\bm{r})$ is also changing with time,
there is an additional term proportional to 
$\dot V(\bm{r})$ on the right side of Eq.~(\ref{sweep}).  
Tan referred to that equation as the {\it dynamic sweep theorem}.

The simplest case of a time-dependent scattering length is a 
sudden change in $a$, for which the change in the energy is given in
Eq.~(\ref{sudden}).  Tan also presented a more general result 
for a scattering length $a(t)$ that changes rapidly enough that 
the contact does not have time to evolve significantly 
from its original value.  If the scattering length $a(t)$ changes 
over a short time interval $T$ from an initial value $a(0)$ to a final value
$a(T)$, the change in the total energy is
\begin{eqnarray}
\Delta E = 
-\frac{\hbar^2}{4\pi m}
\left( \frac{1}{a(T)} - \frac{1}{a(0)} 
+ \sqrt{\frac{8 \hbar}{\pi m}} \int_0^T \!\! \hbox{d}t \int_0^t \!\! \hbox{d}t' 
\sqrt{t-t'} \, \dot b(t) \dot b(t') \right) C,
\nonumber \\
\label{rapid}
\end{eqnarray}
where $C$ is the contact at the initial time $t=0$.
The time interval $T$ 
must be short compared to the time scales for the evolution of the 
system.  It must also be long compared to the time scale  
$m r_0^2/\hbar$ set by the range.

Son and Thompson have also considered rapid changes in the scattering length 
for a system that is initially in the unitary limit~\cite{Son:1002}.
The simplest case is a small-amplitude oscillation of the inverse 
scattering length about the unitary limit:
$a(t)^{-1} = \gamma_0 \cos(\omega t)$.  
If $\omega$ is large compared to the Fermi frequency 
$\hbar k_F^2/2m$, the energy density increases at a rate 
that is proportional to the contact density:
\begin{equation}
\frac{\hbox{d}~}{\hbox{d} t} {\cal E} = 
\frac{\gamma_0^2 \hbar^2}{8 \pi m}
\sqrt{\frac{\hbar \omega}{m}}~{\cal C}.
\label{dEdt}
\end{equation}
The frequency $\omega$  must also be small compared to the 
frequency scale $\hbar/mr_0^2$ set by the range.

Another case of a rapidly varying scattering length
considered in Ref.~\cite{Son:1002}
is a pulse $a(t)^{-1}$ that differs from zero only during a short time 
interval.  The change in the total energy 
is proportional to the contact:
\begin{equation}
\Delta E = 
\frac{\hbar^{5/2}}{4 \pi^2 m^{3/2}}
\left(\int_0^\infty \!\! \hbox{d} \omega~ \sqrt{\omega}~
\left| \tilde \gamma(\omega) \right|^2 \right) C,
\label{DeltaE}
\end{equation}
where $\tilde\gamma(\omega)$ is the Fourier transform of the pulse:
$\tilde\gamma(\omega) = \int_0^T \!\hbox{d}t \exp(i \omega t) a(t)^{-1}$.
This is consistent with the more general result in Eq.~(\ref{rapid}),
which does not require the unitary limit 
before and after the pulse.

\section{Making Contact with Experiment}
\label{sec:experiment}

There are some exciting recent developments in the 
study of the universal relations
using experiments with ultracold atoms.
They involve various measurements of the contact and the 
experimental verification of some of the universal relations.

\subsection{Photoassociation}

The Hulet group at Rice University in Houston
has measured the photoassociation rate 
of a balanced mixture of $^6$Li atoms in the lowest 
two hyperfine spin states~\cite{PSKJH:0505}.
The scattering length was controlled by using the Feshbach resonance 
near 834~G.
They measured the photoassociation rate at various values of 
the scattering length, with $1/k_F a$ ranging from about 
$-1.4$ to about $+5.4$.  
The photoassociation laser can excite the closed-channel molecule
that is responsible for the Feshbach resonance.
The photoassociation rate can therefore be interpreted as a measurement 
of the number of closed-channel molecules.
According to the universal relation in Eq.~(\ref{eq:Nmol-C}),
this is proportional to the contact.
However, at the time of the experiment,
the concept of the contact was still unfamiliar.
The number of closed-channel molecules was expected to be 
proportional to the square of the order parameter 
$|\Delta|^2$~\cite{PSKJH:0505}.
In the BEC limit,  $|\Delta|^2 \propto k_F/a$, 
which has the same dependence on $a$ as the contact in 
Eq.~(\ref{C-BEC}). In the BCS limit, the order parameter 
decreases exponentially with $1/|a|$:
$|\Delta|^2 \propto k_F^2 \exp(-\pi/(k_F |a|))$. 
This behavior is dramatically different from the contact in 
Eq.~(\ref{C-BCS}), which decreases like $a^2$ as $a \to 0^-$.
Nevertheless, the measurements of the photoassociation rate 
were compatible with the assumption 
that it was proportional to $|\Delta|^2$,
even for negative values of $1/k_F a$ as large as $-1.4$.

\begin{figure}[t]
\centerline{\includegraphics*[width=4.in,angle=0]{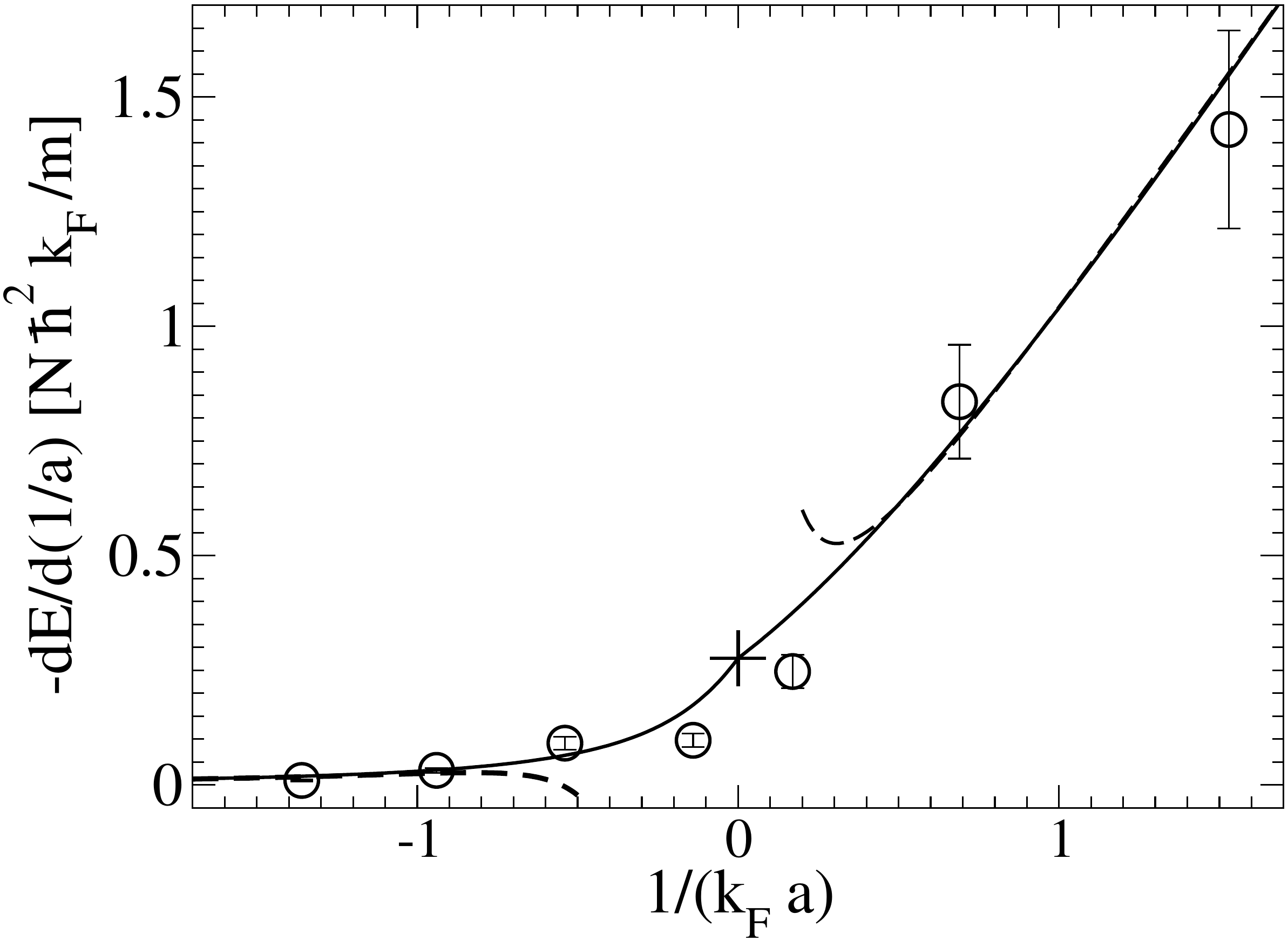}}
  \caption{The derivative of the energy $E$ with respect to $1/a$
    determined from measurements of the photoassociation rate 
    of a trapped gas of $^6$Li atoms, from Ref.~\cite{WTC:0807}.
    The data come from measurements of the number of 
    closed-channel molecules in Ref.~\cite{PSKJH:0505}.
    The solid line is a theoretical prediction 
    using the local density approximation, with the contact density 
    for the homogeneous system obtained by interpolating 
    between the BCS, unitary, and BEC limits.
    The symbol $+$ indicates the prediction for the contact in the 
    unitary limit.
    The dashed lines are extrapolations from the BCS and BEC limits.
}
\label{fig:photoassoc}
\end{figure}
The first analysis of the data from Ref.~\cite{PSKJH:0505}
in terms of the contact was carried out by  
Werner, Tarruell, and Castin~\cite{WTC:0807}.
Their results are shown in Fig.~\ref{fig:photoassoc}.
The contact extracted from the measured number of 
closed-channel molecules was in reasonable agreement
with a theoretical prediction using the local density approximation,
with the contact density for the homogeneous system 
obtained by interpolating between the BCS, unitary, and BEC limits.

\subsection{Static structure factor}

The static and dynamic structure factors 
for systems consisting of ultracold atoms can be probed by using 
Bragg spectroscopy.  Bragg scattering is a two-photon process 
in which an atom absorbs a photon from one laser beam 
and emits a photon into a second laser beam.
The net effect is the transfer of a selected momentum $\hbar \bm{k}$
and a selected energy $\hbar \omega$ to the atom,
where $\bm{k}$ is the difference between the wavevectors 
of the two lasers and $\omega$ is the difference between their frequencies.

\begin{figure}[t]
\centerline{\includegraphics*[width=4.in,angle=0]{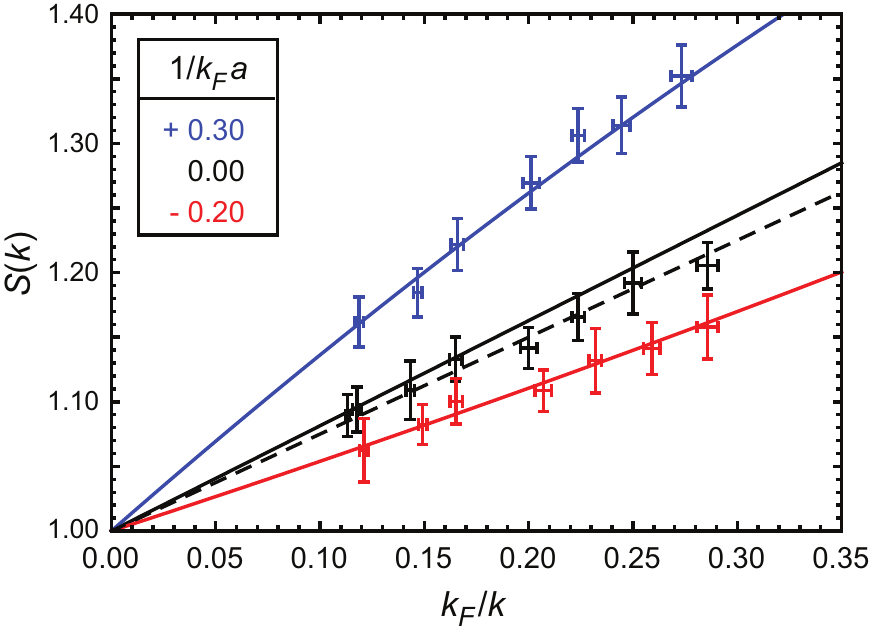}}
  \caption{The structure factor $S(k)$ for a trapped gas of $^6$Li atoms
    as a function of $k_F/k$ for three values of $1/k_F a$, from
    Ref.~\cite{Drummond:1001}. The solid lines are the universal
    predictions from Eq.~\eqref{S12-tail} using the contact 
    obtained from a theoretical calculation.  
    The dashed line is a linear fit to the data for $1/k_F a =0$.
}
\label{fig:structure}
\end{figure}
The Vale group at Swinburne University of Technology 
in Melbourne has used Bragg spectroscopy 
to study the static structure factor $S_{12}(k)$~\cite{VKDJ:0809}.
They used a balanced mixture of $^6$Li atoms in the lowest 
two hyperfine spin states. 
The scattering length was controlled by using the Feshbach resonance 
near 834~G.
In Ref.~\cite{Drummond:1001}, they reported measurements
of the static structure factor $S_{12}(k)$ 
as a function of $a$ for $k = 4.8~k_F$.
These measurements are in good agreement with the 
universal relation for the large-momentum tail in Eq.~(\ref{S12-tail}),
with the contact density for the homogeneous sytem obtained 
by interpolating between the BCS, unitary, and BEC limits.
In Ref.~\cite{Drummond:1001}, they also reported measurements
of $S_{12}(k)$ as a function of $k$ for $1/k_F a = -0.2$, 0, and $+0.3$,
which are shown in Fig.~\ref{fig:structure}.  
The measurements are linear in $k_F/k$, 
as predicted by the universal relation in Eq.~(\ref{S12-tail}).
The slope is predicted to be proportional to the contact $C$.
The contact for their trapped system at zero temperature
was calculated in Ref.~\cite{Drummond:1001} 
using the local density approximation and a below-threshold 
Gaussian fluctuation theory for the homogeneous system. 
For $1/k_F a = -0.2$ and $+0.3$, the slope agrees well 
the universal relation.
In the unitary limit $1/k_F a = 0$, the slope 
is smaller than predicted.
The discrepancy could be attributed to the effects of 
nonzero temperature.

\subsection{Comparing measurements of the contact}
\label{sec:measurecontact}

The Jin group at JILA in Boulder has measured the contact $C$ 
for a trapped gas of atoms using three independent methods~\cite{Jin1002}.
They used a balanced mixture of $^{40}K$ atoms in the two lowest 
hyperfine spin states at a temperature of about $0.1~T_F$.
The scattering length was controlled by using the Feshbach resonance 
near 201~G.
\begin{figure}[t]
\centerline{\includegraphics*[width=4.in,angle=0]{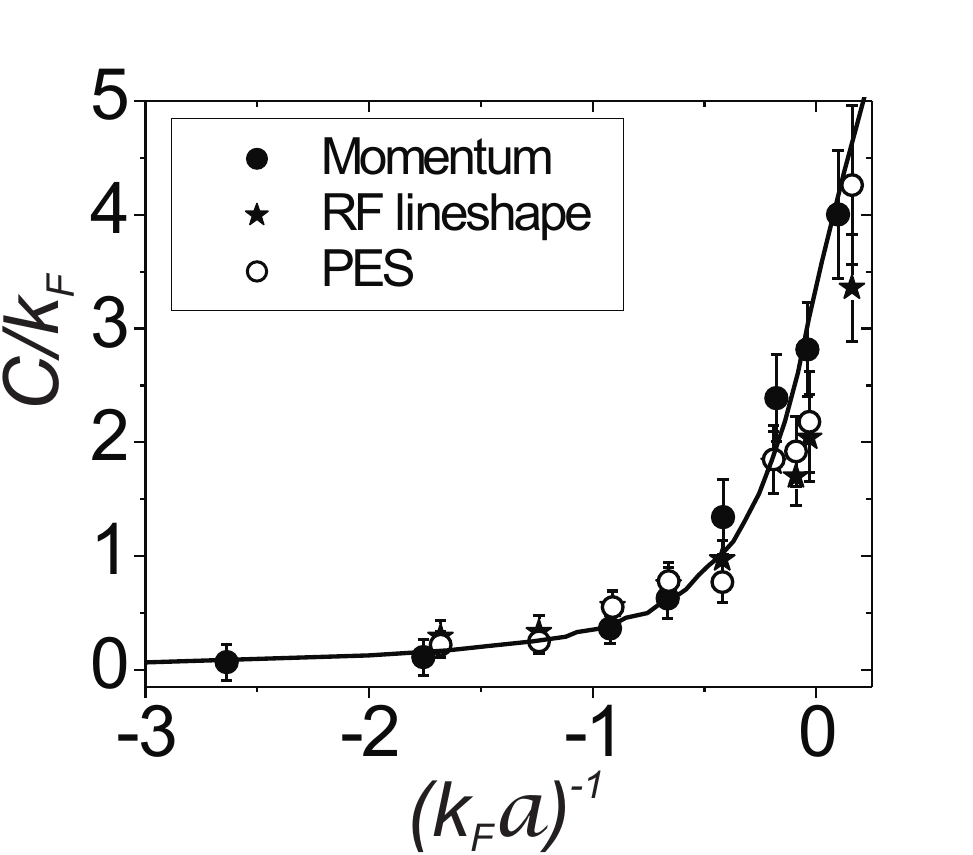}}
  \caption{Three measurements of the dimensionless contact $C/k_F$
  for a trapped gas of $^{40}K$ atoms
  as a function of $1/k_F a$, from Ref.~\cite{Jin1002}. 
  The Fermi wavenumber $k_F$ is defined by the Fermi energy for the trapped
  system: $E_F = \hbar^2 k_F^2/2m$.
  Two of the data sets are from the tail of the momentum distribution
  measured directly by ballistic expansion (solid dots)  
  and indirectly by photoemission spectrometry (open dots).  
  The third data set is from the high-frequency tail 
  of the rf lineshape (stars).
}
\label{fig:measurecontact}
\end{figure}

The first method for measuring $C$ used the Tan relation in 
Eq.~(\ref{tails}) for the high-momentum tail of the momentum
distribution $n_2(k)$.
The interactions between the trapped atoms were turned off by changing the
magnetic field to 209~G where the scattering length vanishes.
The trapping potential was then turned off, 
and the momentum distribution was
measured from the ballistic expansion of the cloud of atoms.
The contact $C$ is the large-momentum limit of $k^4 n_2(k)$.
It was measured for values of $1/k_F a$ ranging from about 
$-2.7$ to about $+0.2$.

The second method for measuring the contact used the universal relation in 
Eq.~(\ref{eq:Gamma-high2}) for the high-frequency tail of the 
radio-frequency (rf) transition rate.  
The rf signal was used to transfer atoms from state 2 
into a third spin state 3 for which the pair scattering 
length $a_{13}$ is not large.
The rf transition rate $\Gamma(\omega)$ was determined by 
measuring the number of atoms transferred.  The contact was then 
determined from the behavior of $\Gamma(\omega)$ at large $\omega$.
It was measured for values of $1/k_F a$ ranging from about 
$-1.7$ to about $+0.2$.

The third method for measuring $C$ also used the Tan relation in 
Eq.~(\ref{tails}), but the tail of the momentum
distribution was determined by photoemission spectroscopy (PES).  
This involves using momentum-resolved rf spectroscopy 
to measure the distribution $n_2(k,\omega)$ 
of the momentum and energy of atoms in state 2, and then integrating over 
$\omega$ to determine the momentum distribution $n_2(k)$.
The contact $C$ is the large-momentum limit 
of $k^4 n_2(k)$.  It was measured for the same values of $1/k_F a$
as the second method.

The three sets of measurements of the contact by the Jin group  
\cite{Jin1002} are shown in Figure~\ref{fig:measurecontact}.  
The results from the three methods are all 
consistent.  They lie close to the theoretical prediction 
of Ref.~\cite{WTC:0807}, which was based on the local density approximation, 
with the contact density for the homogeneous system obtained 
by interpolating between the BEC, unitary, and BCS limits.
These results provide direct experimental verification of the role 
of the contact in large-momentum and high-frequency tails 
for a many-body system of fermions with a large scattering length.

\subsection{Tests of the thermodynamic Tan relations}

The adiabatic relation in Eq.~(\ref{adiabatic})
and the virial theorem in Eq.~(\ref{virial}) relate different 
contributions to the total energy $E$ to the contact $C$.
The adiabatic relation expresses a derivative of $(T+U)+V$ 
in terms of the contact. 
The virial theorem expresses the combination $(T+U)-V$ 
in terms of the contact.  Thus measurements of $T+U$, $V$,
and $C$ provide two nontrivial tests of the Tan relations
for the thermodynamic properties of the system. 

\begin{figure}[t]
\centerline{\includegraphics*[width=4.in,angle=0]{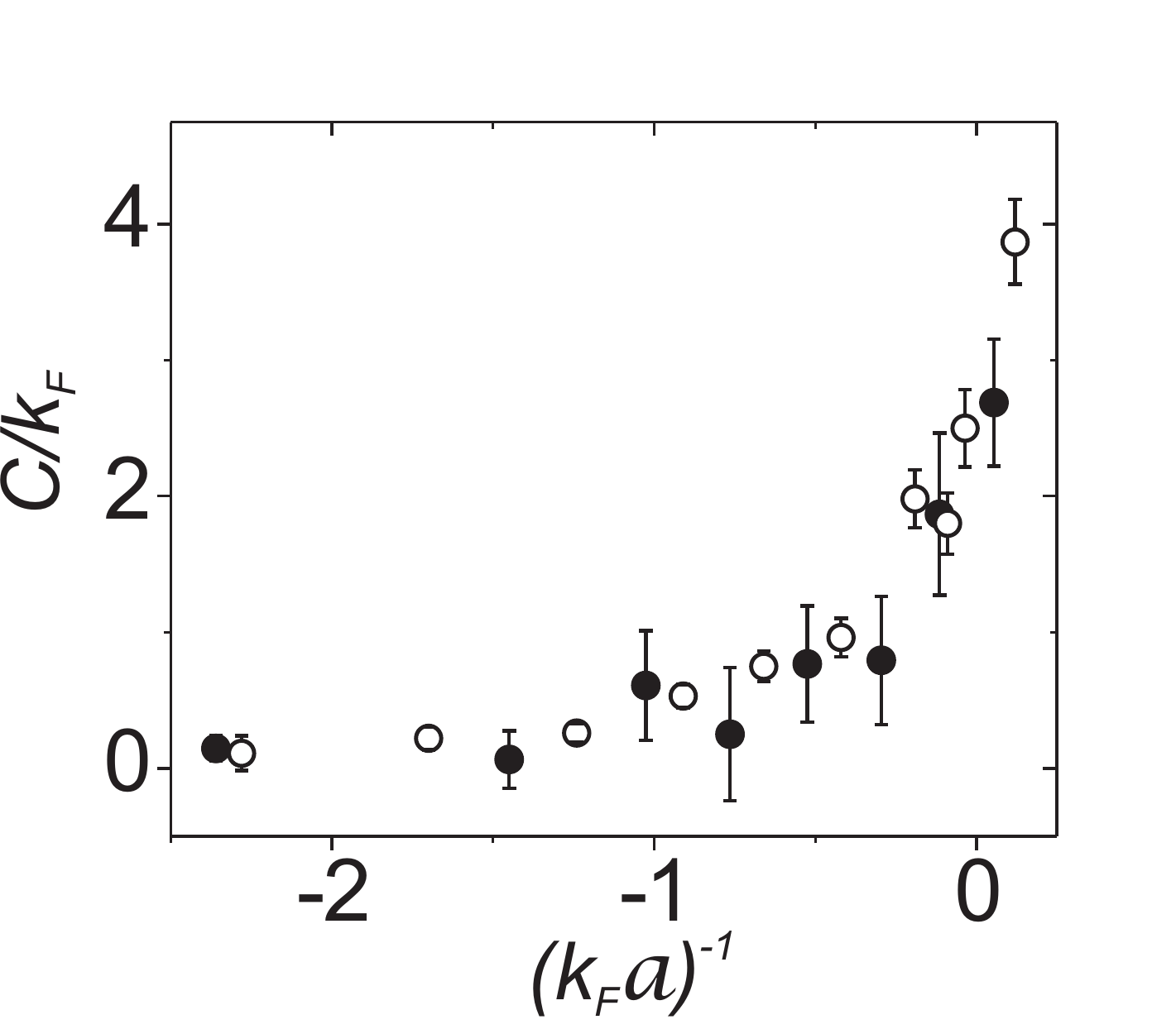}}
  \caption{Test of the adiabatic relation
  in a trapped gas of $^{40}K$ atoms, from Ref.~\cite{Jin1002}.
  The dimensionless contact $C/k_F$ as a function of $1/k_F a$
  determined from the derivative of the energy $E$ 
  with respect to $1/a$ (solid dots) is compared 
  to the measurements using photoemission spectrometry (open dots).
  The Fermi wavenumber $k_F$ is defined by the Fermi energy for the trapped
  system: $E_F = \hbar^2 k_F^2/2m$.
}
\label{fig:testuniversal}
\end{figure}
The Jin group at JILA in Boulder has tested these Tan relations by measuring
$T+U$ and $V$  for the same system of $^{40}K$ atoms 
for which they measured the contact $C$~\cite{Jin1002}, as 
described in Section~\ref{sec:measurecontact}.
They measured the external potential energy $V$ by imaging the spatial
distribution of the cloud of atoms, which was trapped in a harmonic potential.
They measured the combination $T+U$, which can be called the release energy,
by turning off the trapping potential 
and observing the resulting expansion of the cloud.
They measured $T+U$ and $V$ at values of $1/k_F a$ ranging from about 
$-3$ to about $+0.3$.
For the contact $C$, they used their measurements from
photoemission spectroscopy described in Section~\ref{sec:measurecontact}.
They found good agreement between the two sides of the 
adiabatic relation in Eq.~(\ref{adiabatic})
as shown in Figure~\ref{fig:testuniversal}.
They also found that the two sides of the 
virial theorem in Eq.~(\ref{virial}) agreed to within the errors, 
which were roughly 1\% of the Fermi energy.
These results provide direct experimental verification of the role 
of the contact in the thermodynamics of 
a many-body system of fermions with a large scattering length.

\subsection{Contact density near unitarity}
\label{sec:contactdensity}

The contact density for the homogeneous gas has been determined 
by the Salomon group at \'Ecole Normale Sup\'erieure in Paris~\cite{Salomon:1004}.
They used an imbalanced mixture of  $^6$Li atoms 
in the lowest two hyperfine spin states at a magnetic field near 834~G.
They determined the equation of state for the homogeneous gas 
by measuring the number densities of the two spin states
in a harmonic trapping potential integrated 
over the two tranverse dimensions. 
Their result for the numerical constant in Eq.~(\ref{C-unitary})
was $\zeta = 0.93(5)$.  This is consistent with the value in 
Eq.~(\ref{zeta}) obtained from diffusion Monte Carlo calculations
of the density-density correlator.

\section{Derivations of Universal Relations}
\label{sec:deriv-univ-results}

In this section, we give an overview of various derivations of the
universal relations.  We begin by describing briefly the novel methods
used in the original derivations of the Tan relations. 
We then describe briefly various other approaches that have been used 
to rederive the Tan relations.  Finally, we describe 
in more detail how universal relations can be derived 
using quantum field theory
methods involving renormalization and the operator product expansion.

\subsection{Preliminaries}
\label{sec:prelim}

The scattering amplitude for S-wave atom-atom scattering can be
written as
\begin{equation}
f(k)= \frac{1}{k \cot\delta_0(k) -i k} ,
\label{f-k}
\end{equation}
where $k$ is the relative wavenumber and $\delta_0(k)$ is the 
S-wave phase shift.  If the interactions have a finite range, 
the low-energy expansion of the phase shift can be
expressed as a power series in $k^2$:
\beq
k\cot \delta_0(k) = -1/a + \mbox{$\frac{1}{2}$} r_s k^2 + \ldots  ,
\label{kcot}
\eeq
where $r_s$ is the effective range.
The coefficient of $(k^2)^n$ has dimensions (length)$^{2n-1}$.
Generically, all these coefficients are comparable to the 
range raised to the appropriate power.
By fine tuning the interactions, the scattering length can be made 
much larger than the range.
This fine-tuning leads to universal properties that depend on the
interactions only through the scattering length.  

The universality for large scattering length 
reflects a well-behaved {\it zero-range limit}, 
in which all the terms on the right side of Eq.~(\ref{kcot}) go to zero 
except the leading term.  The phase shift reduces in this limit to
\begin{eqnarray}
k \cot \delta_0(k) = - 1/a .
\label{delta0:ZRM}
\end{eqnarray}
The model in which the phase shift has this simple form
up to arbitrarily large momentum is called the {\it Zero-Range Model}.
The universal properties of a general model with large scattering length
are realized in the Zero-Range Model in a particularly simple form,
because $a$ is the only length
scale that arises from interactions.  The price that must be paid 
is that the zero-range limit leads to divergences in some observables.
It also leads to singularities in intermediate steps of 
the derivations of universal relations. 
An illustration is provided by the energy relation in Eq.~(\ref{energy}).
The kinetic energy $T$ and the interaction energy $U$ are separately 
ultraviolet divergent, but the divergences cancel in the sum $T+U$.
The singularities associated with the zero-range limit
can be regularized by backing off from the zero-range limit
or by some equivalent device.

One way to represent the Zero-Range Model is in terms of the Schr\"odinger
equation for noninteracting particles with peculiar boundary conditions.
The Schr\"odinger wavefunction for $N_1$ fermions in state 1 and
$N_2$ fermions in state 2 is a complex function
$\Psi(\bm{r}_1,\ldots,\bm{r}_{N_1};\bm{r}_1',\ldots,\bm{r}_{N_1}')$
that is totally antisymmetric in the first $N_1$ positions and  
totally antisymmetric in the last $N_2$ positions. 
The proper normalization of the wavefunction is
\begin{eqnarray}
\frac{1}{N_1! N_2!}
\int \!\! \hbox{d}^3r_1 \ldots \int \!\! \hbox{d}^3r_{N_1}
\int \!\! \hbox{d}^3r_1' \ldots \int \!\! \hbox{d}^3r_{N_2}'
\left| \Psi(\bm{r}_1, ..., {\bf r}_{N_1}; 
          \bm{r}_1', ..., {\bf r}_{N_2}') \right|^2 = 1. 
\nonumber
\\
\label{eq:Psi-1}
\end{eqnarray}
In the zero-range limit, this wavefunction diverges when the positions
$\bm{r}_i$ and $\bm{r}_j'$ of any pair of fermions with different spins
coincide.  Its behavior when $\bm{r}_1$ and $\bm{r}_1'$ are nearly equal is
\begin{eqnarray}
\Psi(\bm{R} \mbox{$+\frac12$} \bm{r},\bm{r}_2, ..., {\bf r}_{N_1};
    \bm{R} \mbox{$-\frac12$} \bm{r},\bm{r}_2', ..., {\bf r}_{N_2}') 
\nonumber\\
\longrightarrow \phi(r)~ 
\Phi(\bm{r}_2, ..., {\bf r}_{N_1}; \bm{r}_2', ..., {\bf r}_{N_2}';\bm{R}) , 
\label{eq:Bethe-Peierls}
\end{eqnarray}
where $\Phi$ is a smooth function of $\bm{R}$ 
and $\phi(r)$ is the zero-energy scattering wavefunction 
for two particles interacting through a large scattering length:
\begin{equation}
\phi(r) = \frac{1}{r} - \frac{1}{a} .
  \label{eq:zeroscat}
\end{equation}
The Fourier transform of this wavefunction is 
\begin{equation}
\tilde \phi(k) = \frac{4 \pi}{k^2} - \frac{(2 \pi)^3}{a} \delta^3(\bm{k}) .
\label{eq:zeroscatFT}
\end{equation}
The Schr\"odinger equation for 
interacting particles with a large scattering length reduces 
in the zero-range limit to the Schr\"odinger equation for  
non-interacting particles with the wavefunction constrained to 
satisfy the Bethe-Peierls boundary conditions in 
Eq.~(\ref{eq:Bethe-Peierls}).

\subsection{Tan's derivations}
\label{sec:tans-derivation}

Tan derived many of his universal relations by using generalized 
functions, or distributions, to deal with the singularities 
associated with the 
zero-range limit~\cite{Tan:0505,Tan:0508}.  He introduced
distributions $\Lambda(\bm{k})$ and $L(\bm{k})$
whose values at finite $\bm{k}$ are
\begin{subequations}
\begin{eqnarray}
\Lambda(\bm{k}) &=& 1, \qquad |\bm{k}| < \infty,
\label{Lambda1}
\\
L(\bm{k}) &=& 0, \qquad |\bm{k}| < \infty,
\label{L1}
\end{eqnarray}
\end{subequations}
and which have the following integrals over all $\bm{k}$ :
\begin{subequations}
\begin{eqnarray}
\int \!\! \frac{\hbox{d}^3k}{(2 \pi)^3} \frac{1}{k^2} \Lambda(\bm{k}) &=& 0,
\label{Lambda2}
\\
\int \!\! \frac{\hbox{d}^3k}{(2 \pi)^3} \frac{1}{k^2} L(\bm{k}) &=& 1.
\label{L2}
\end{eqnarray}
\end{subequations}
Using these properties, it is easy to verify that the Fourier 
transform of the zero-energy scattering wavefunction
in Eq.~(\ref{eq:zeroscatFT}) satisfies
\begin{equation}
\int \!\! \frac{\hbox{d}^3k}{(2 \pi)^3} 
\left[ \Lambda(\bm{k}) + \frac{1}{4 \pi a}  L(\bm{k}) \right]
\tilde \phi(\bm{k}) = 0.
\label{eq:zeroscatBP}
\end{equation}
This property allows the Bethe-Peierls boundary condition in 
Eq.~(\ref{eq:Bethe-Peierls}) to be expressed as an equality 
rather than as a limit:
\begin{eqnarray}
&&\int \!\! \frac{\hbox{d}^3k}{(2 \pi)^3} 
\left[ \Lambda(\bm{k}) + \frac{1}{4 \pi a}  L(\bm{k}) \right]
\int \!\! \hbox{d}^3r~e^{-i \bm{k} \cdot \bm{r}}
\nonumber \\
&& \hspace{1.25cm} \times 
\Psi(\bm{R} \mbox{$+\frac12$} \bm{r},\bm{r}_2, ..., {\bf r}_{N_1};
    \bm{R} \mbox{$-\frac12$} \bm{r},\bm{r}_2', ..., {\bf r}_{N_2}') 
    = 0 .
\label{eq:PsiBP}
\end{eqnarray}

In addition to using 
$\Lambda(\bm{k})$ and $L(\bm{k})$ to impose the 
Bethe-Peierls boundary conditions, Tan used 
the distribution $\Lambda(\bm{k})$ 
to regularize the interaction term in the Hamiltonian.
He derived many of his universal relations by simple
manipulations involving these distributions.
They lead to an expression for the contact of the form
\begin{equation}
C = 
\frac12 \sum_\sigma \int \!\! \frac{\hbox{d}^3k}{(2 \pi)^3} 
L(\bm{k}) k^2 n_\sigma(\bm{k}) .
\label{eq:C-L}
\end{equation}
By the properties of the distribution $L(\bm{k})$
in Eqs.~(\ref{L1}) and (\ref{L2}), the integral extracts the 
coefficient of $1/k^4$ in the high-momentum tail of $n_\sigma(\bm{k})$.

\subsection{Other derivations}
\label{sec:quant-mech-meth}

\subsubsection{Braaten, Kang, and Platter}

Braaten and Platter~\cite{Braaten:2008uh} 
used quantum field theory methods to rederive all 
the Tan relations in Section~\ref{sec:tan-relations}
except for the relation in 
Eq.~(\ref{sudden}) for the sudden change in the scattering length. 
They used the formulation of the Zero-Range Model as a 
local quantum field theory, 
as described later in Section~\ref{sec:zero-range-model}.
The singularities associated with the zero-range limit were 
regularized by imposing an ultraviolet cutoff on the momenta 
of virtual particles.  
The contact density was identified as the expectation value 
of a local operator.
The universal relations were derived using 
renormalization, as described in Section~\ref{sec:renormalization}, 
and the operator product expansion,
as described in Sections~\ref{sec:short-dist-OPE}
and \ref{sec:short-time-OPE}.

Braaten, Kang, and Platter~\cite{Braaten:2008bi}
used quantum field theory methods to derive universal relations
for the {\it Resonance Model}, in which the S-wave scattering 
phase shift is given by
\begin{eqnarray}
k \cot \delta_0(k) = 
- \left( \lambda + \frac{g^2}{k^2 - \nu} \right)^{-1} .
\label{delta0:ResM}
\end{eqnarray}
The Resonance Model is a 2-channel model, in which the 
states in the 2-atom sector consist of a point-like molecule 
as well as the usual 2-atom scattering states.
It provides a natural model for a Feshbach resonance.
The scattering length $a = \lambda - g^2/\nu$ agrees with 
the expression in Eq.~(\ref{eq:aFeshbach}) if we set
$\lambda = a_\textrm{bg}$, 
$\nu = - m \mu_\textrm{mol} (B - B_0)/\hbar^2$,
and $g^2 = 1/R_*$, where $R_*$ is given in Eq.~(\ref{eq:R*}).
Braaten, Kang, and Platter found that in the various 
universal relations that correspond to the Tan relations,
the contact density is replaced by various linear combinations 
of the expectation values of three local operators.  
If the expression for $k \cot \delta_0(k)$
in Eq.~(\ref{delta0:ResM}) is well-approximated by $-1/a$
for all wavenumbers from 0 up to the scale set by the system,
the expectation values of the three local operators
must coincide in order for the universal relations 
to reduce to those of the Zero-Range Model.
For a broad Feshbach resonance,
which is defined by $|a_\textrm{bg}| \gg R_*$,
this requires only that $|a| \gg r_0$ and $k r_0 \ll 1$,
where $r_0$ is the range of interactions in the absence 
of the Feshbach resonance.
For a narrow Feshbach resonance,
which is defined by $|a_\textrm{bg}| \ll R_*$,
this requires also that $k R_* \ll 1$.

\subsubsection{Werner, Tarruell, and Castin}

Werner~\cite{Werner:0803} derived the virial theorem 
in Eq.~(\ref{virial}) independently.
The right side of Eq.~(\ref{virial}) was expressed not
in terms of the contact $C$ but in terms of the derivative of the energy 
that appears in the adiabatic relation in Eq.~(\ref{adiabatic}). 

Werner, Tarruell, and Castin~\cite{WTC:0807} rederived 
the tail of the momentum distribution in Eq.~(\ref{tails}), 
the adiabatic relation, and the 
density-density correlator at short distances in Eq.~\eqref{nncor}.
The singularities associated with the zero-range limit 
were regularized by using a lattice model in which the fermions 
occupy the sites of a 3-dimensional cubic lattice 
whose spacing $b$ approaches 0.
Werner, Tarruell and Castin also used a two-channel model to derive
the universal relation for the number of closed-channel molecules
in Eq.~(\ref{eq:Nmol-C}). 

Werner and Castin~\cite{WC:1001} subsequently presented 
a much more thorough treatment of the universal relations for the 
Zero-Range Model and for the lattice model.
In addition to the tail of the momentum distribution, 
the adiabatic relation, and the 
density-density correlator at short distances,
they rederived the energy relation in Eq.~(\ref{energy})
and the adiabatic sweep theorem, which is the generalization
of Eq.~(\ref{sweep})
that allows for a time-dependent external potential.
They showed that for a system in thermal equilibrium 
(which includes the ground state as a limiting case), 
the contact is an increasing function of $1/a$:
\begin{equation}
\frac{\hbox{d}C~~}{\hbox{d} a^{-1}} > 0 . 
\label{eq:ddC}
\end{equation}
This inequality holds whether the derivative is evaluated at fixed entropy 
or at fixed temperature.
The monotonic increase of the contact density with $1/a$
is illustrated in Figs.~\ref{fig:balanced} and 
\ref{fig:imbalanced} for the cases of a homogeneous gas at zero temperature
that is balanced and strongly imbalanced, respectively.
Because the inequality in Eq.~(\ref{eq:ddC}) does not hold in general,
it should not be regarded as a universal relation.

Werner and Castin also generalized 
the universal relations to various other systems~\cite{WC:1001}.  
They considered the effects of a nonzero range 
for the interaction potential.  They derived the generalizations 
of the universal relations to two spatial dimensions, 
which had been considered previously by Tan~\cite{Tan:0505615}.
They also derived universal relations 
for two types of fermions with unequal masses $m_1 < m_2$, 
for identical bosons,
and for mixtures of fermions and bosons with various masses.
In some cases, including identical bosons 
and two types of fermions whose mass ratio $m_2/m_1$
exceeds the critical value 13.7,
the universal relations are complicated by the Efimov effect~\cite{WC:1001}.

\subsubsection{Zhang and Leggett}

Zhang and Leggett~\cite{Zhang:0809} rederived the adiabatic relation
in Eq.~(\ref{adiabatic}) and the pressure relation in Eq.~(\ref{pressure}).
They used a nonlocal quantum field theory 
with field operators $\psi_1(\bm{r})$ and $\psi_2(\bm{r})$
and with a short-range interaction potential $U(\bm{r})$.
They restricted their attention to a homogeneous many-body system 
in equilibrium.
The equal-time 2-particle correlation function 
was expressed as a sum over eigenstates of the operator 
$\psi_2({\bf r}_2)\psi_1(\bm{r}_1)$:
\begin{equation}
\langle \psi^\dagger_1(\bm{r}_1) \psi^\dagger_2(\bm{r}_2)
        \psi_2({\bf r}_2)\psi_1(\bm{r}_1)\rangle = 
\sum_i \nu_i~\phi^{(i)}(\bm{r}_1,\bm{r}_2)^*~\phi^{(i)}(\bm{r}_1,\bm{r}_2) ,
\label{eq:<>sum}
\end{equation}
where $\nu_i$ is the average number of pairs with different spins
in the eigenstate $i$ and the eigenfunctions $\phi^{(i)}(\bm{r}_1,\bm{r}_2)$ 
are normalized accordingly in a large volume $V$.
The Bethe-Peierls boundary conditions together with translation 
invariance imply that the limiting behavior of these eigenfunctions 
as the separation $r = |\bm{r}_1 - \bm{r}_2|$ goes to zero is
\begin{equation}
\phi^{(i)}(\bm{R} \mbox{$+\frac12$} \bm{r}, 
          \bm{R} \mbox{$-\frac12$} \bm{r})
\longrightarrow C^{(i)} ~
e^{i \bm{P}^{(i)} \cdot \bm{R}}~\phi(r) ,
\label{phiphi}
\end{equation}
where the normalization constant $C^{(i)}$ 
and the momentum vector $\bm{P}^{(i)}$ depend on the eigenstate $i$
and $\phi(r)$ is the zero-energy scattering wavefunction
defined in Eq.~\eqref{eq:zeroscat}.  The integral of the 
correlation function weighted by an arbitrary short-distance function 
$s(|\bm{r}_1-\bm{r}_2|)$ therefore reduces to:
\begin{eqnarray}
  \int \!\! \hbox{d}^3 r_1 \int \!\! \hbox{d}^3 r_2~s(|\bm{r}_1-\bm{r}_2|)
  \langle \psi^\dagger_1(\bm{r}_1) \psi^\dagger_2(\bm{r}_2)
        \psi_2({\bf r}_2)\psi_1(\bm{r}_1)\rangle
\nonumber \\
=\frac{1}{4 \pi} V {\cal C}
\int_0^\infty \!\! \hbox{d}r~r^2 s(r) |\phi(r)|^2 ,
\label{eq:ints<>}
\end{eqnarray}
where ${\cal C}$ is the contact density, which is given by
\begin{equation}
{\cal C} = 16 \pi^2 \sum_i \nu_i | C^{(i)} |^2 .
\label{eq:Csum}
\end{equation}
Zhang and Leggett emphasized that the thermodynamics is universal 
and is completely determined by the contact density ${\cal C}$.
Zhang and Leggett also used a two-channel model to derive a 
universal relation
for the number of closed-channel molecules 
that can be reduced to Eq.~(\ref{eq:Nmol-C})~\cite{Zhang:0809}. 

Zhang and Leggett derived a factorization formula for the
interaction energy density~\cite{Zhang:0809}
that separates the dependence on $a$ and thermodynamic variables, 
such as the temperature and number densities,
from the dependence on the short-distance parameters 
that determine the shape of the interaction potential $U(r)$.
The interaction energy is the special case of Eq.~(\ref{eq:ints<>}) 
in which the short-range function $s(r)$
is the interaction potential $U(r)$.
The interaction energy density can be expressed as
\begin{equation}
{\cal U} = \frac{1}{4 \pi}~{\cal C}
\int_0^\infty \!\! \hbox{d}r~r^2 U(r) |\phi(r)|^2 .
\label{eq:U-C}
\end{equation}
All the dependence on the thermodynamic variables
and on the large scattering length $a$
resides in the contact density ${\cal C}$.

\subsubsection{Combescot, Alzetto, and Leyronas}

Combescot, Alzetto, and Leyronas~\cite{CAL:0901} rederived the 
tail of the momentum distribution in Eq.~(\ref{tails}) 
and the energy relation in Eq.~(\ref{energy}).  
They used the Schr\"odinger formalism in the coordinate representation 
to describe a system consisting of $N_1 + N_2$ fermions.
The singularities associated with the zero-range limit were 
regularized by imposing a cutoff $|\bm{r}_i - \bm{r}_j'| > r_0$ 
on the separations of the two types of fermions.  They expressed
the contact in terms of the Fourier transform of the function $\Phi$ 
defined by the Bethe-Peierls boundary conditions in 
Eq.~(\ref{eq:Bethe-Peierls}).
By reverting to the coordinate representation,
the contact density can be expressed as
\begin{eqnarray}
{\cal C}(\bm{R}) = \frac{16 \pi^2}{(N_1-1)! (N_2-1)!}
\int \!\! \hbox{d}^3r_2 \ldots \int \!\! \hbox{d}^3r_{N_1}
\int \!\! \hbox{d}^3r_2' \ldots \int \!\! \hbox{d}^3r_{N_2}'
\nonumber
\\
\times \left| \Phi(\bm{r}_2, ..., {\bf r}_{N_1}; 
          \bm{r}_2', ..., {\bf r}_{N_2}';\bm{R}) \right|^2 , 
\label{eq:C-BP}
\end{eqnarray}
provided the wavefunction $\Psi$ is properly normalized 
as in Eq.~(\ref{eq:Psi-1}).
This expression for the contact density 
was originally derived by Tan~\cite{Tan:0505}.

Combescot, Alzetto, and Leyronas generalized the 
tail of the momentum distribution and the energy relation 
to various other systems~\cite{CAL:0901}.
They generalized them to two spatial dimensions, 
which had been considered previously by Tan~\cite{Tan:0505615}.
They generalized them to two types of fermions with
different masses $m_1 < m_2$.
They did not however consider the complications associated
with the Efimov effect when the mass ratio $m_2/m_1$
exceeds the critical value 13.7~\cite{WC:1001}.

\subsection{Quantum field theory derivations}
\label{sec:QFTderive}

We proceed to explain how universal relations can be derived concisely
using the methods of quantum field theory.

\subsubsection{Zero-Range Model}
\label{sec:zero-range-model}

A quantum field theory that describe atoms with two spin states 
must have fundamental quantum fields $\psi_\sigma(\bm{r})$, 
$\sigma = 1, 2$. The Hamiltonian operator for a
local quantum field theory can be expressed as the integral over space of a
Hamiltonian density operator: $H =\int \hbox{d}^3R \, {\cal H}$.
If the atoms are in an external potential $V(\bm{r})$, 
the Hamiltonian density operator is the sum of a kinetic term ${\cal T}$, 
an interaction term ${\cal U}$, and an external potential term ${\cal V}$: 
\begin{equation}
{\cal H} = {\cal T} + {\cal U} + {\cal  V} .
\label{Hdens}
\end{equation}
In the quantum field theory formulation of the Zero-Range Model, 
the three terms in the Hamiltonian density operator are
\begin{subequations}
\begin{eqnarray}
{\cal T} &=& 
\sum_\sigma \frac{1}{2m} 
        \nabla \psi_\sigma^\dagger \cdot \nabla \psi_\sigma^{(\Lambda)}(\bm{R}),
\label{T}
\\
{\cal U} &=& 
\frac{g_0(\Lambda)}{m} \psi_1^\dagger \psi_2^\dagger \psi_2 \psi_1^{(\Lambda)}(\bm{R}),
\label{U}
\\
{\cal V} &=& V(\bm{R}) \sum_\sigma \psi_\sigma^\dagger \psi_\sigma(\bm{R}).
\label{V}
\end{eqnarray}
\label{H}
\end{subequations}
For simplicity, we have set $\hbar = 1$. 
The superscripts $(\Lambda)$ on the
operators in Eqs.~(\ref{T}) and (\ref{U}) indicate that their matrix elements
are ultraviolet divergent and an ultraviolet cutoff is required to make them
well defined.  For the ultraviolet cutoff, we impose an upper limit
$|\bm{k}|<\Lambda$ on the momenta of virtual particles.  
In the limit $\Lambda \to \infty$, the Hamiltonian density 
in Eq.~(\ref{Hdens}) describes atoms with the phase shift given by 
Eq.~(\ref{delta0:ZRM}) provided we take the bare coupling constant to be
\begin{equation}
g_0(\Lambda) = \frac{4 \pi a}{1 - 2 a \Lambda/\pi} .
\label{g2}
\end{equation}

In Ref.~\cite{Braaten:2008uh}, Braaten and Platter identified the 
operator that measures the contact density
in the quantum field theory formulation of the Zero-Range Model.
It is convenient to introduce the diatom field operator $\Phi$
defined by
\begin{equation}
\Phi(\bm{R}) = g_0(\Lambda) \, \psi_2 \psi_1^{(\Lambda)}(\bm{R}) .
\label{Phi-psi}
\end{equation}
This operator annihilates a pair of atoms at the point $\bm{R}$.
The superscript $(\Lambda)$ on the operator $\psi_1 \psi_2$ 
indicates that its matrix elements are ultraviolet divergent.
Their dependence on $\Lambda$ is exactly compensated by the prefactor
$g_0(\Lambda)$, so $\Phi$ is an ultraviolet finite operator.
The contact density operator is $\Phi^\dagger \Phi$~\cite{Braaten:2008uh}.
This is just the interaction energy density operator multiplied 
by a constant that depends on the ultraviolet cutoff: 
$\Phi^\dagger \Phi = m g_0 {\cal U}$.
The contact is obtained by taking the expectation value 
of the contact density operator and integrating over space:
\begin{equation}
C = \int \!\! \hbox{d}^3R~ \langle \Phi^\dagger \Phi(\bm{R}) \rangle .
\label{C-Phi}
\end{equation}

\subsubsection{Renormalization}
\label{sec:renormalization}

Several of the Tan relations follow very simply from the 
renormalization of the Zero-Range Model~\cite{Braaten:2008uh}.
The renormalization condition in Eq.~(\ref{g2}) 
implies that the bare coupling constant $g_0(\Lambda)$ satisfies
\begin{equation}
g_0(\Lambda) = 
\left( \frac{1}{4 \pi a} - \frac{\Lambda}{2 \pi^2} \right)
g_0^2(\Lambda) .
\label{g02}
\end{equation}
Its derivative with respect to $a$ is
\begin{equation}
\frac{\hbox{d}\ }{\hbox{d} a} g_0(\Lambda) = 
\frac{1}{4 \pi a^2}g_0^2(\Lambda) .
\label{dg0da}
\end{equation}
The energy relation in Eq.~(\ref{energy}) and
the adiabatic relation in Eq.~\eqref{adiabatic} follow simply from these
properties of the bare coupling constant:
\begin{description}
\item
{\bf Energy relation}.
The kinetic and interaction terms ${\cal T}$ and ${\cal U}$ 
in the Hamiltonian density operator are given in Eqs.~(\ref{T}) and (\ref{U}).
After inserting the expression in Eq.~(\ref{g02}) 
for the bare coupling constant $g_0$ into ${\cal U}$,
the sum of ${\cal T}$ and ${\cal U}$ can be expressed as 
the sum of two finite operators:
\begin{eqnarray}
{\cal T} + {\cal U} = 
\Big( \sum_\sigma \frac{1}{2m} 
\nabla \psi_\sigma^\dagger \cdot \nabla \psi_\sigma^{(\Lambda)}
- \frac{ \Lambda}{2 \pi^2 m} 
\Phi^\dagger \Phi
\Big)
+ \frac{1}{4 \pi m a } \Phi^\dagger \Phi .
\label{H-finite}
\end{eqnarray}
By taking the expectation value of both sides of Eq.~(\ref{H-finite}),
integrating over space, and using the
expression for $C$ in Eq.~(\ref{C-Phi}).
we obtain the energy relation in Eq.~(\ref{energy}). \\

\item
{\bf Adiabatic relation}.
According to the Feynman-Hellman theorem, the rate of change in the energy 
due to a change in the scattering length can be expressed in the form
\begin{equation}
\left(\frac{\hbox{d} E}{\hbox{d}a}\right)_S = 
\int \!\! \hbox{d}^3R \left\langle \frac{\partial{\cal H}}{\partial a}\right \rangle  .
\label{E-H}
\end{equation}
The Hamiltonian density ${\cal H}$ depends on $a$ 
only through the factor of $g_0$ in ${\cal U}$.
Using the derivative of the bare coupling constant in Eq.~(\ref{dg0da}), 
we obtain the derivative of the Hamiltonian density:
\begin{equation}
\frac{\partial{\cal H}}{\partial a} =
 \frac{1}{4 \pi m a^2} \Phi^\dagger \Phi  .
\label{H-a}
\end{equation}
By inserting this into Eq.~(\ref{E-H})
and using the expression for $C$ in Eq.~(\ref{C-Phi}),
we obtain the adiabatic relation in Eq.~\eqref{adiabatic}.
\end{description}

The virial theorem in Eq.~\eqref{virial}
and the pressure relation in Eq.~\eqref{pressure}
can be derived by combining renormalization with dimensional analysis.
For the purposes of dimensional analysis, 
we can regard $\hbar=1$ and $m$ simply as conversion constants that 
allow any dimensionful quantity to be expressed as a length raised to 
an appropriate power.
\begin{description}
\item
{\bf Virial theorem}. 
For a system in a harmonic trapping potential,
the only parameters that an energy eigenvalue can depend on are 
the scattering length $a$ and the angular frequency $\omega$.
The combinations with dimensions of length are $a$ and 
$(m \omega)^{-1/2}$.
Since an energy eigenvalue $E$ 
has dimensions $1/(m \, \hbox{length}^2)$,
the constraint of dimensional analysis can be expressed as the 
requirement that a differential operator that counts the factors 
of length gives $-2$ when acting on 
$E = \int \hbox{d}^3R \, \langle {\cal H} \rangle$:  
\begin{equation}
\left( a \frac{\partial \ }{\partial a} 
- 2 \omega \frac{\partial\ }{\partial \omega} \right)
\int \!\! \hbox{d}^3R~ \langle {\cal H} \rangle = 
-2 E  .
\label{virial-2}
\end{equation}
Using the Feynman-Hellman theorem, 
this equation can be written 
\begin{equation}
\int \!\! \hbox{d}^3R 
\left( \frac{1}{4 \pi m a} \langle\Phi^\dagger \Phi\rangle
- 4  \langle{\cal V}\rangle  \right) = 
-2 E  .
\label{virial-3}
\end{equation}
Using the expression for $C$ in Eq.~(\ref{C-Phi}),
we obtain the virial theorem in Eq.~\eqref{virial}.\\

\item
{\bf Pressure relation}.
For a homogeneous system,
the only variables that the free energy density 
${\cal F} = {\cal E} - T {\cal S}$
can depend on are the scattering length $a$, 
the temperature $T$, and the number densities $n_i$.
The combinations with dimensions of length are $a$,
$(m k_B T)^{-1/2}$, and $n_i^{-1/3}$.
Since ${\cal F}$ has dimensions $1/(m \, \hbox{length}^5)$,
the constraint of dimensional analysis can be expressed as the 
requirement that a differential operator that counts the factors 
of length gives $-5$ when acting on ${\cal F}$:  
\begin{equation}
\left( a \frac{\partial~}{\partial a} 
- 2 T \frac{\partial~}{\partial T}  
- 3 n_1 \frac{\partial~}{\partial n_1} 
- 3 n_2 \frac{\partial~}{\partial n_2} \right) {\cal F} = 
-5 {\cal F}  .
\label{pressure-2}
\end{equation}
Using the adiabatic relation in Eq.~(\ref{adiabatic}), 
this can be written
\begin{equation}
\frac{1}{4 \pi m a} {\cal C}
+ 2 T {\cal S} - 3 \mu_1 n_1 - 3 \mu_2 n_2 = 
-5 {\cal F}  ,
\label{pressure-3}
\end{equation}
where $\mu_i$ is the chemical potential for the spin state $i$.
The pressure relation in Eq.~\eqref{pressure} then follows from the
thermodynamic identity
${\cal F} = - {\cal P} + \mu_1 n_1 + \mu_2 n_2$.

\end{description}

\subsubsection{The operator product expansion}
\label{sec:OPE}

Many universal relations are most concisely derived using the 
{\it operator product expansion} (OPE) of quantum field theory.  
The OPE was invented independently in 1969 by three giants 
of theoretical physics:  Leo Kadanoff~\cite{Kadanoff:1969}, 
Alexander Polyakov~\cite{Polyakov:1969},
and Ken Wilson~\cite{Wilson:1969}.  It is an expansion 
for the product of local operators at nearby points 
in terms of local operators at the same point:
\begin{equation}
O_A(\bm{R} + \mbox{$\frac{1}{2}$}\bm{r})~
O_B(\bm{R} - \mbox{$\frac{1}{2}$} \bm{r})
= \sum_C f_{A,B}^C(\bm{r})~O_C(\bm{R}).
\label{OPE-sd}
\end{equation}
The {\it Wilson coefficients} $f_{A,B}^C (\bm{r})$ are ordinary 
functions of the separation vector $\bm{r}$. 
The local operators $O_C(\bm{R})$ 
include some that can be obtained by Taylor-expanding the
operators on the left side in powers of $\bm{r}$,
but they also include additional operators that 
take into account effects from quantum fluctuations.
The Wilson coefficients for these operators are not necessarily 
analytic functions of the vector $\bm{r}$, and they can 
even diverge as $\bm{r} \to 0$.  
One particularly simple local operator is the unit operator $I$,
whose expectation value in any state is 1.
The sum over $C$ in Eq.~(\ref{OPE-sd})
can be extended to include all local operators 
if we allow Wilson coefficients that are 0.

The OPE is a natural tool for generating universal relations,
because it is an operator identity. 
By taking the expectation value of both sides of the OPE in 
Eq.~(\ref{OPE-sd}) in some state of the system, 
we get an expression for the correlator of the 
operators $O_A$ and $O_B$ in terms of the 
expectation values of local operators in that state.
Since this expansion holds for any state of the system, it is  
a universal relation.  

A local operator $O_C(\bm{R})$ is assigned the
{\it scaling dimension} $d_C$ if the correlation function of 
$O_C$ and its hermitian conjugate at points separated by $r$ 
decreases asymptotically as $1/r^{2 d_C}$ at small $r$.
The unit operator $I$ is assigned scaling dimension 0.
In a weakly-interacting theory, the scaling dimensions 
can be obtained simply by dimensional analysis.
In a strongly-interacting theory, they can be significantly different.  
The difference between the scaling dimension and 
its value in the corresponding
noninteracting theory is called the {\it anomalous dimension}.
At very small $\bm{r}$, the leading behavior of a nontrivial 
Wilson coefficient is determined by the scaling dimensions
of the operators:
\begin{equation}
f_{A,B}^C(\bm{r}) \sim r^{d_C - d_A - d_B}.
\label{f-r}
\end{equation}
In the OPE in Eq.~(\ref{OPE-sd}), the Wilson coefficients
of higher dimension operators go to 0 more rapidly as $r \to 0$.
A Wilson coefficient can be suppressed by a further power
of $r$ is there is an explicitly broken symmetry which, 
if exact, would require $f_{A,B}^C(\bm{r})$ to vanish.
The extra suppression factor is the dimensionless 
combination of $r$ and the symmetry breaking parameter.

The technical assumptions required to prove the OPE 
have been discussed by Wilson and Zimmerman~\cite{Wilson:1972}.
The OPE can be expressed more precisely 
as an asymptotic expansion in the separation $r = |\bm{r}|$.  
The OPE in Eq.~(\ref{OPE-sd}) can be organized into an 
expansion in powers of $r$ by
expanding the Wilson coefficients as Laurent series in $r$.
The OPE is an asymptotic expansion if  
for any power $p$, there are only finitely many terms 
that go to zero more slowly than $r^p$.
The scaling behavior of the Wilson coefficients in Eq.~(\ref{f-r})
guarantees that the OPE is an asymptotic expansion 
provided every local operator $O_C(\bm{R})$ has a positive
scaling dimension $d_C>0$
and there are only finitely many local operators 
with scaling dimension $d_C < d$ for any positive number $d$.
These conditions are satisfied in the Zero-Range Model, 
the Resonance Model, and other renormalizable local 
quantum field theories that are relevant to cold atoms.

An illustration of the operator product expansion 
with anomalous dimensions is
provided by the Ising Model in 2 dimensions.  The exact solution by
Lars Onsager in 1944~\cite{Onsager:1944} implies that correlation
functions in the continuum limit have scaling behavior 
with anomalous dimensions.  For
example, the leading term in the correlation function for two spin
operators $\sigma$ as their separation $\bm{r}$ goes to 0 has
the form
\begin{equation}
\left \langle \sigma(\bm{R} + \mbox{$\frac{1}{2}$}\bm{r})~
\sigma(\bm{R} - \mbox{$\frac{1}{2}$} \bm{r}) \right \rangle
\longrightarrow \frac{A}{ |\bm{r}|^{1/4}},
\label{Ising-correlator}
\end{equation}
where $A$ is a constant that does not depend on the state of the system.
This correlator is singular as $r \to 0$.
The power law behavior suggests that the system becomes scale invariant 
at short distances.  The fractional power of $r$ indicates that the spin
operator has an anomalous dimension.

Kadanoff generalized the result for the correlator in 
Eq.~(\ref{Ising-correlator}) to an operator relation~\cite{Kadanoff:1969}:
\begin{equation}
\sigma(\bm{R} + \mbox{$\frac{1}{2}$} \bm{r})~
\sigma(\bm{R} - \mbox{$\frac{1}{2}$} \bm{r}) =
\frac{A}{|\bm{r}|^{1/4}} I
+ B  |\bm{r}|^{3/4} \varepsilon(\bm{R}) 
+ \ldots ,
\label{Ising-OPE}
\end{equation}
where $I$ is the identity operator,
$\varepsilon(\bm{R})$ is the energy fluctuation operator,
and $B$ is another constant.
The infinitely many terms that are not shown explicitly 
in Eq.~(\ref{Ising-OPE}) go to 0 more rapidly than $r^{3/4}$ as $r \to 0$. 
The short-distance tail of the correlator in 
Eq.~(\ref{Ising-correlator}) can be obtained simply by taking 
the expectation value of Eq.~(\ref{Ising-OPE}).  
Kadanoff showed that the critical exponents of the Ising model, 
such as the exponent $\frac14$ of $1/|\bm{r}|$ in Eq.~(\ref{Ising-correlator}),
could be deduced simply from the knowledge of which operator products have
singular Wilson coefficients.
From the OPE in Eq.~(\ref{Ising-OPE}), 
we can deduce that the spin operator has scaling dimension $\frac18$ 
and the energy fluctuation operator has scaling dimension 1.

The OPE in Eq.~(\ref{OPE-sd}) can be described more precisely as a
{\it short-distance operator product expansion},
because it involves operators at the same time with small spatial separation.
It can be generalized to a {\it short-time operator product expansion},
in which the operators also have a small separation in time:
\begin{eqnarray}
{\cal O}_A(\bm{R} \mbox{$+\frac12$} \bm{r},T \mbox{$+\frac12$} t)~
{\cal O}_B(\bm{R} \mbox{$-\frac12$} \bm{r},T \mbox{$-\frac12$} t)
= \sum_C f^{C}_{A,B}(\bm{r},t)~{\cal O}_C (\bm{R},T) .
\nonumber
\\ 
\label{OPE-gen}
\end{eqnarray}
The Wilson coefficients $f^{C}_{A,B}(\bm{r},t)$ are functions of 
the separation vector $\bm{r}$ and the time interval $t$. 
The short-time OPE is more subtle than the short-distance OPE
because of the possibility that a correlator can have 
oscillatory behavior in $t$ as $t \to 0$~\cite{Wilson:1972}.
The possibility of oscillatory behavior is avoided in the 
Euclidean version of the quantum field theory that corresponds
to analytic continuation of the time $t$ to 
Euclidean time: $t \to -i \tau$.
Thus the short-time OPE in Eq.~(\ref{OPE-gen}) can be expressed 
most rigorously as an asymptotic expansion in $(\bm{r},-i \tau)$,
where $\tau$ is the Euclidean time separation obtained by the 
analytic continuation $t \to -i \tau$.
In a Galilean-invariant theory, the appropriate scaling 
of $(\bm{r},-i \tau)$
is such that $\tau$ scales in the same way as $|\bm{r}|^2$.

\subsubsection{Short-distance operator product expansion}
\label{sec:short-dist-OPE}

Universal relations for fermions with large scattering length 
can be derived from operator product expansions in the 
Zero-Range Model defined in Section~\ref{sec:zero-range-model}.
In this model, the scattering length $a$ is the only length
scale that arises from interactions.  At distances much smaller than 
$|a|$, the model is scale invariant with nontrivial scaling dimensions.
In the unitary limit $a \to \pm\infty$, 
the model is not only scale invariant at all distances
but also conformally invariant~\cite{Son:2005rv}. 
If $a$ is finite, we can regard $1/a$ as the symmetry-breaking 
parameter associated with the broken conformal symmetry.

The fundamental field operators $\psi_1$ and $\psi_2$  
of the Zero-Range Model have the same
scaling dimensions as in a noninteracting theory.
However there are composite operators with anomalous scaling dimensions.
The scaling dimension of an operator
${\cal O}_C$ can be deduced from its propagator at large 
momentum $\bm{k}$, which in a Galilean-invariant theory
scales as $k^{2 d_C - 5}$.  Since the propagator for 
$\psi_1$ or $\psi_2$ is simply $(\omega - k^2/2m)^{-1}$, 
these fields have scaling dimensions $\frac32$.  
The scaling dimension of the number density operator 
$\psi_\sigma^\dagger \psi_\sigma$ 
is twice that of $\psi_\sigma$, which is 3.
If there were no interactions, the scaling dimension 
of the composite operator $\psi_1 \psi_2$, 
or equivalently the diatom field operator $\Phi$ defined in 
Eq.~(\ref{Phi-psi}), would also be 3. 
However the propagator for $\Phi$ is
\begin{eqnarray}
\int \!\! \hbox{d}t~e^{i \omega t} 
   \int \!\! \hbox{d}^3r~e^{-i \bm{k} \cdot \bm{r}}~
\langle \Phi(\bm{r},t)~\Phi^\dagger(\bm{0},0) \rangle
= \frac{-i 4 \pi m}{-1/a + \sqrt{-m(\omega-k^2/4m)}}.
\nonumber \\
\label{prop-Phi}
\end{eqnarray}
We have dropped an additive constant that is independent of 
$\omega$ and $\bm{k}$, which could be removed by renormalization.
Since the propagator in Eq.~(\ref{prop-Phi}) scales as $k^{-1}$ at large $k$,
$\Phi$ has scaling dimension 2 and therefore anomalous dimension $-1$.
The scaling dimension of the contact density operator 
$\Phi^\dagger \Phi$ is twice that of $\Phi$, which is 4.

The short-distance OPE can be used to derive the Tan relation 
for the tail of the momentum distribution in Eq.~(\ref{tails}).
The momentum distribution $n_{\sigma}(\bm{k})$ for atoms 
in the spin state $\sigma$ can be expressed as
\begin{equation}
n_{\sigma} (\bm{k}) = 
\int \!\! \hbox{d}^3 R \!\! \int \!\! \hbox{d}^3r~e^{-i \bm{k} \cdot \bm{r}} 
\langle \psi_\sigma^\dagger(\bm{R} \mbox{$-\frac12$} \bm{r})~
\psi_\sigma(\bm{R} \mbox{$+\frac12$} \bm{r}) \rangle .
\label{rho-psi}
\end{equation}
The behavior at large $\bm{k}$ is dominated by the small-$\bm{r}$ 
region of the integral.
We can therefore apply the OPE to the product of the operators
$\psi_\sigma^\dagger$ and $\psi_\sigma$.
As shown in Ref.~\cite{Braaten:2008uh},
the leading terms in the OPE are
\begin{eqnarray}
\psi_\sigma^\dagger(\bm{R} \mbox{$-\frac12$} \bm{r})~ 
\psi_\sigma(\bm{R} \mbox{$+\frac12$} \bm{r}) &=& 
\psi_\sigma^\dagger \psi_\sigma(\bm{R})
\nonumber
\\
&& + \mbox{$\frac{1}{2}$} \bm{r} \cdot
\left[ \psi_\sigma^\dagger \nabla \psi_\sigma(\bm{R}) 
- \nabla \psi_\sigma^\dagger \psi_\sigma(\bm{R}) \right]
\nonumber
\\
&& - \frac{r}{8 \pi} \Phi^\dagger \Phi(\bm{R})
+ \ldots  .
\label{OPE-ZRM}
\end{eqnarray}
We have written explicitly all terms whose Wilson coefficients go to
zero more slowly than $r^2$ as $r \to 0$.
The first two terms on the right
side of the OPE in Eq.~(\ref{OPE-ZRM}) can be obtained by multiplying the
Taylor expansions of the two operators.  The third term arises from quantum
fluctuations involving pairs of atoms with small separations.  
That its Wilson coefficient is proportional to $r$ can be predicted 
from the scaling dimensions of the operators using Eq.~(\ref{f-r}).
The coefficient of $r$ can be calculated using diagrammatic methods 
described in Ref.~\cite{Braaten:2008bi}.
Note that this Wilson coefficient is not an analytic function of the vector 
$\bm{r}= (x,y,z)$ at $\bm{r} = 0$, because it is proportional 
to $r = \sqrt{x^2 + y^2 + z^2}$.
The expectation value of the OPE in Eq.~(\ref{OPE-ZRM}) 
can be expressed as
\begin{eqnarray}
\langle \psi_\sigma^\dagger(\bm{R} \mbox{$-\frac12$} \bm{r})~ 
\psi_\sigma(\bm{R} \mbox{$+\frac12$} \bm{r}) \rangle &=& 
n_\sigma(\bm{R})
+ i \bm{r} \cdot \bm{{\cal P}}_\sigma(\bm{R})
- \frac{r}{8 \pi} {\cal C}(\bm{R})
+ \ldots  ,
\nonumber \\
\label{<OPE>-ZRM}
\end{eqnarray}
where $\bm{{\cal P}}_\sigma$ is the momentum density of atoms 
in the state $\sigma$.  This form of the OPE was first 
written down by Tan~\cite{Tan:0505}.

Upon inserting the OPE in Eq.~(\ref{OPE-ZRM}) into Eq.~(\ref{rho-psi}),
the first two terms give a delta function in $\bm{k}$ 
and the gradient of such a delta function.
They correspond to contributions to $n_{\sigma} (\bm{k})$
that decrease at large $\bm{k}$ faster than any power of $k$.
In the third term, the Fourier transform of the Wilson coefficient 
at nonzero values of $\bm{k}$ can be obtained from the identity
\begin{equation}
\int \!\! \hbox{d}^3r~  e^{-i \bm{k} \cdot \bm{r}} r  =
- \frac{8 \pi}{k^4} ,
\label{r-ft}
\end{equation}
which can be derived by differentiating the Fourier transform 
of a $1/r$ potential.
This term gives a power-law tail in the momentum distribution:
\begin{equation}
n_{\sigma} (\bm{k}) \longrightarrow 
\frac{1}{k^4}
\int \!\! \hbox{d}^3 R \, \langle \Phi^\dagger \Phi(\bm{R}) \rangle .
\label{rho-k4}
\end{equation}
Comparing with Eq.~(\ref{tails}), we obtain the expression 
 in Eq.~(\ref{C-Phi}) for the contact $C$ in the Zero-Range Model.
This verifies that the contact density operator is $\Phi^\dagger \Phi$, 
where $\Phi$ is the diatom field operator defined in Eq.~(\ref{Phi-psi}).

The short-distance OPE can be used to derive the
Tan relation in Eq.~(\ref{nncor}) for the density-density correlator 
at short distances~\cite{Braaten:2008uh}.
The OPE for the number density operators
$\psi_1^\dagger\psi_1$ and $\psi_2^\dagger\psi_2$ includes a term
whose Wilson coefficient is singular as $\bm{r} \to 0$:
\begin{equation}
\psi_1^\dagger\psi_1(\bm{R} \mbox{$+\frac12$}\bm{r})~
\psi_2^\dagger\psi_2(\bm{R} \mbox{$-\frac12$}\bm{r})
= \frac{1}{16 \pi^2} \left( \frac{1}{r^2} - \frac{2}{a r} \right)
\Phi^\dagger \Phi(\bm{R}) + \ldots .
\label{OPE:n1n2}
\end{equation}
All the other terms in the OPE are regular at $\bm{r} = 0$.
Taking the expectation value of both sides, we get the 
Tan relation in Eq.~(\ref{nncor}).

The OPE in Eq.~(\ref{OPE:n1n2}) can also be used to derive the 
universal relation for the static structure factor $S_{12}(q)$
in Eq.~(\ref{S12-tail}).  That structure factor can be expressed as
\begin{equation}
S_{12}(q) = 
\int \!\! \hbox{d}^3 R \!\! \int \!\! \hbox{d}^3r~  e^{-i \bm{q} \cdot \bm{r}} 
\langle \psi_1^\dagger \psi_1(\bm{R} \mbox{$-\frac12$} \bm{r})
\psi_2^\dagger \psi_2(\bm{R} \mbox{$+\frac12$} \bm{r}) \rangle .
\label{S12-def}
\end{equation}
According to the universal relation in Eq.~(\ref{S12-tail}),
the high momentum tail has terms proportional to $1/q$ and $1/q^2$.
They come from the singular terms proportional to $1/r^2$ and $1/r$
in the OPE in Eq.~(\ref{OPE:n1n2}).

\subsubsection{Short-time operator product expansion}
\label{sec:short-time-OPE}

Other universal relations can be derived using the short-time 
operator product expansion in Eq.~\eqref{OPE-gen}.
We will illustrate the use of the short-time OPE by 
showing how it can be used to derive the universal relations
for radio-frequency (rf) spectroscopy 
that were presented in Section~\ref{sec:rf}. 

The rf signal that causes
a transition of an atom in spin state 2 into an atom 
in spin state 3 corresponds
to the action of an operator $\psi^\dagger_3\psi_2(\bm{r},t)$. 
The inclusive rate $\Gamma(\omega)$ for the production of 
atoms in state 3 can be expressed as
\begin{eqnarray}
\Gamma(\omega) &=& \Omega^2 ~{\rm Im} \, i \!\!
\int \!\! \hbox{d}t~e^{i (\omega + i \epsilon) t} 
\int \!\! \hbox{d}^3R \int \!\! \hbox{d}^3r
\nonumber\\
&& \times 
\langle \hbox{T} \psi_2^\dagger \psi_3(\bm{R} \mbox{$+\frac12$} \bm{r},t)~ 
 \psi_3^\dagger \psi_2(\bm{R} \mbox{$-\frac12$} \bm{r},0) \rangle.
\label{I-psipsi}
\end{eqnarray}
The symbol T in the matrix element implies that the product of operators is
time ordered.  For a large frequency $\omega$,
the integrals are dominated by small Euclidean time intervals 
$t = - i \tau$ and by small separations $\bm{r}$.  
We can therefore apply the short-time OPE in Eq.~(\ref{OPE-gen}) to the
product of operators $\psi_2^\dagger \psi_3$ and $\psi_3^\dagger \psi_2$.
The Wilson coefficients of the leading one-body operator
$\psi_2^\dagger \psi_2$ and the leading two-body operator 
$\Phi^\dagger \Phi$ were determined in Ref.~\cite{Braaten:2010dv}:
\begin{eqnarray}
&& \int \!\! \hbox{d}t~e^{i \omega t} \int \!\! \hbox{d}^3r~
 \psi_2^\dagger \psi_3(\bm{R} \mbox{$+\frac12$} \bm{r},t)~ 
 \psi_3^\dagger \psi_2(\bm{R} \mbox{$-\frac12$} \bm{r},0)
\nonumber \\
&&   \hspace{0.5cm} 
= \frac{i}{\omega}~\psi_2^\dagger \psi_2(\bm{R})
\nonumber \\
&&   \hspace{1.0cm} 
+ \frac{i (a_{12}^{-1} - a_{13}^{-1}) [a_{12}^{-1} - \sqrt{-m \omega}\,]} 
    {4 \pi m \omega^2 [a_{13}^{-1} - \sqrt{-m \omega}\,]}~\Phi^\dagger \Phi(\bm{R})
+ \ldots.
\label{psipsi-OPE}
\end{eqnarray}
Upon inserting the OPE in Eq.~(\ref{psipsi-OPE}) into Eq.~(\ref{I-psipsi}),
the $\psi_2^\dagger \psi_2$ term gives a delta function in $\omega$.
This corresponds to a contribution that decreases at large $\omega$ 
faster than any power.   The leading contribution to the 
high-frequency tail comes from the $\Phi^\dagger \Phi$ term:
\begin{eqnarray}
\Gamma(\omega) \longrightarrow
\frac{\Omega^2 (a_{13}^{-1} - a_{12}^{-1})^2}
    {4 \pi m \omega^2} \hbox{Im} 
\left( \frac{-1}{a_{13}^{-1} - \sqrt{-m (\omega + i \epsilon)}} \right)
\int \!\! \hbox{d}^3 R \, \langle \Phi^\dagger \Phi(\bm{R}) \rangle .
\nonumber \\
\label{tail-omega123}
\end{eqnarray}
This reduces to the universal relation for the high-frequency tail in 
Eq.~(\ref{eq:Gamma-high}).

Sum rules for the rf transition rate $\Gamma(\omega)$,
such as those in Eqs.~(\ref{eq:Gamma-sum0}) and (\ref{eq:Gamma-sum1}),
can be derived by expressing the integral along the real $\omega$ 
axis as a contour integral in the complex $\omega$ plane 
that wraps around the real axis.
If we allow for a general weighting function $f(\omega)$, 
the sum rule becomes
\begin{eqnarray}
\int_{-\infty}^{\infty} \!\!\hbox{d} \omega~f(\omega) \, \Gamma(\omega) &=& 
\Omega^2 ~\oint \!\! \frac{\hbox{d} \omega}{2 \pi}~f(\omega)
\int \!\! \hbox{d}t~e^{i \omega t} 
\int \!\! \hbox{d}^3R \int \!\! \hbox{d}^3r
\nonumber\\
&& \times 
\langle \hbox{T} \psi_2^\dagger \psi_3(\bm{R} \mbox{$+\frac12$} \bm{r},t)~ 
 \psi_3^\dagger \psi_2(\bm{R} \mbox{$-\frac12$} \bm{r},0) \rangle.
\label{sumrule-f}
\end{eqnarray}
If that contour is deformed into a circle whose radius approaches 
infinity, then $\omega$ has a large imaginary part along 
most of the contour.  This justifies the use of the 
short-time OPE in Eq.~(\ref{psipsi-OPE}).
The sum rule can then be derived by evaluating the contour integral
along the circle at infinity for each of the Wilson coefficients.
For $f(\omega) = 1$, the only nonzero contribution is from the
$\psi_2^\dagger \psi_2$ term and the sum rule reduces to 
Eq.~(\ref{eq:Gamma-sum0}).
For $f(\omega) = \omega$, the only nonzero contribution is from the
$\Phi^\dagger \Phi$ term and the sum rule reduces to 
Eq.~(\ref{eq:Gamma-sum1}).
The weighting function $f(\omega)$ does not need to be a polynomial 
in $\omega$.  The Lorentzian function 
$f(\omega) = [(\omega - \omega_0)^2 + \gamma^2]^{-1}$
gives a family of sum rules with two adjustable parameters 
that is less sensitive to range corrections than that in 
Eq.~(\ref{eq:Gamma-sum1})~\cite{Braaten:2010dv}.

There are other universal relations that can be
derived using the short-time OPE.
One example is the high-frequency behavior of the structure factor
$S_{12}(\omega,q)$,
such as the universal relation in Eq.~(\ref{S12-tail2}).
It can be derived from the short-time OPE for the number density operators
$\psi_1^\dagger \psi_1$ and $\psi_2^\dagger \psi_2$.

\begin{acknowledgement}
This research was supported in part by a joint grant from the 
Army Research Office and the Air Force Office of Scientific Research.
I would like to acknowledge useful comments by Lucas Platter,
Shina Tan, Edward Taylor, Felix Werner, and Shizhong Zhang.
\end{acknowledgement}

\end{document}